\documentclass[a4paper]{article}

\usepackage{graphicx}
\usepackage{tikz}
\usepackage{amsmath}
\usepackage{amssymb}
\usepackage{algpseudocode}
\usepackage{algorithm}
\usepackage{hyperref}
\usepackage{changes}
\usepackage{mathtools}

\newcommand{\bs}[1]{\boldsymbol{#1}}
\newcommand{\p}{\partial}
\newcommand{\D}{\mathrm{d}}

\newcommand{\Grad}{\nabla}
\newcommand{\Div}[1]{\nabla \cdot {#1}}
\newcommand{\Lap}{\Delta}
\newcommand{\SPH}[1]{\left\langle {#1} \right\rangle}

\makeatletter
\newsavebox{\@brx}
\newcommand{\llangle}[1][]{\savebox{\@brx}{\(\m@th{#1\langle}\)}%
  \mathopen{\copy\@brx\kern-0.5\wd\@brx\usebox{\@brx}}}
\newcommand{\rrangle}[1][]{\savebox{\@brx}{\(\m@th{#1\rangle}\)}%
  \mathclose{\copy\@brx\kern-0.5\wd\@brx\usebox{\@brx}}}
\makeatother

\newcommand{\SPHint}[1]{\llangle {#1} \rrangle}

\ifx\elsarticle\undefined
\renewcommand{\added}[1]{#1}
\renewcommand{\deleted}[1]{#1}
\fi

\begin{document}

\title{Boundary conditions for SPH through energy conservation}

\newcommand{\abstracttext}{
Dealing with boundary conditions in Smoothed Particle Hydrodynamics (SPH) poses significant difficulties, indeed being one of the SPHERIC Grand Challenges.
In particular, wall boundary conditions have been pivotal in SPH model development since it evolved from astrophysics to more generic fluid dynamics simulations.
Despite considerable attention from researchers and numerous publications dedicated to formulating and assessing wall boundary conditions, few of them have addressed the crucial aspect of energy conservation.
This work introduces a novel boundary condition designed with energy conservation as a primary consideration, effectively extending the unconditional stability of SPH to problems involving wall boundary conditions.
\added{%
The result is formulated within the framework of the Boundary Integrals technique.
}
The proposal is tested on a number of cases: normal impact against a wall, adiabatic oscillations of a piston, dam break\added{%
, and the water landing of a spacecraft.
}
}

\ifx\elsarticle\undefined


\author{Cercos-Pita, Jose Luis\\
    CoreDigital department, CoreMarine AS\\
    \texttt{<jlc@core-marine.com>}
\and
    Duque, Daniel \\
    CEHINAV Research Group, ETS Ingenieros Navales,\\
    Universidad Politécnica de Madrid
\and
    Calderon-Sanchez, Javier \\
    CEHINAV Research Group, ETS Ingenieros Navales,\\
    Universidad Politécnica de Madrid
\and
    Merino-Alonso, Pablo Eleazar\\
    CoreDigital department, CoreMarine AS
}

\else


\author[1]{Jose Luis Cercos-Pita
        }
\author[2]{Daniel Duque\corref{cor1}%
        }
\ead{daniel.duque@upm.es}
\author[1,2]{  Pablo Eleazar Merino-Alonso }
\author[2]{ Javier Calderon-Sanchez }
\cortext[cor1]{Corresponding author}
\affiliation[1]{
organization={CoreDigital department, CoreMarine AS},
postcode={4014},
city={Stavanger},
country={Norway}}
\affiliation[2]{
organization={CEHINAV Research Group, ETS Ingenieros Navales, Universidad Politécnica de Madrid},
city={Madrid},
postcode={28040},
country={Spain}}

\begin{abstract}
\abstracttext
\end{abstract}


\fi

\maketitle

\ifx\elsarticle\undefined
\begin{abstract}
\abstracttext
\end{abstract}
\fi

\section{Introduction}

As of today, two of the most fundamental challenges of the Smoothed Particle Hydrodynamics (SPH) methods are boundary conditions (BC)\deleted{,} and stability.

The importance of BC is recognized in the GC\#2 SPHERIC Grand Challenge, named precisely \textit{Boundary conditions} \cite{vacondio2021grand}. At variance with other methods, particle methods present difficulties whenever boundaries are present. Since particles move, it is not trivial to properly describe common boundary types, such as inlet, outlet, or solid wall. This work focuses on the latter, where the issue of impenetrability is paramount. It is also important to translate the usual BC (Neumann, Dirichlet,\ldots) conditions for the particles, an issue that is also challenging, especially in SPH, due to the large support of the kernel functions --- this implies that particles not too close to the walls may still be directly affected by them.

Many proposals have been put forward by SPH researchers in order to prescribe boundary conditions, with impenetrable walls playing a preeminent role.
They may be grouped into three categories \cite{cercospita_thesis_2016}: Boundary Forces (BF), Fluid Extensions (FE), and Boundary Integrals (BI).
These will be discussed in detail in Section \ref{ss:sph:boundaries}, although a brief summary will be sketched in the following.

BF is clearly the more na\"ive and direct approach, based on a set of markers that interact with the fluid particles, exerting a force on them
\cite{Monaghan_kajtar_cpc_2009, crespo2007b}.

FE approaches are conceptually more complex than BF: they consist of refilling the support that is truncated when a particle is close to the boundary. They were historically introduced \added{as early as 1970 by Johnson \textit{et al}} \cite{johnson1970interface}, probably because they are a natural step beyond BF.
The difference in FE methods stems from the details on how to fill the complementary domain, which implies defining both, the placement of the fictitious particles across the boundary, and the field values on them.
Depending on the placement of the particles, three main FE techniques can be distinguished:
\begin{enumerate}
\item Ghost particles \cite{johnson1970interface}. The fluid particles are mirrored with respect to the wall.
\item Dummy particles \cite{lee2008}. A set of fictitious particles is populated in the complementary domain before hand.
\item Local templates \cite{fourtakas2019lust}. Each particle transports a sort of template of fictitious particles that are enabled when they approach the boundary.
\end{enumerate}

Regarding the field values, a number of methodologies have been described and discussed in the past.
Particularly interesting are formulations focusing on consistency, like the ones discussed in Refs. \cite{8th_SPHERIC_Merino_etal, MaciaetalPTP, Marrone2011, antuono2023clone}

The last boundary condition that has been proposed, BI,
is probably the natural evolution of the FE, replacing the missing volume by a surface integral, through Gauss' theorem.
This boundary condition was drafted in Refs.~\cite{campbell1989, kulasegaram_2004, deleffe_etal_spheric09}, but was not really considered an alternative until the first consistent formulation was proposed on Ref. \cite{ferrand_etal_2012}, and analyzed in Ref. \cite{Maciaetal_PTP_2012}.

Regarding stability, its importance is made clear in the GC\#1 SPHERIC Grand Challenge, named \textit{Convergence, consistency and stability} \cite{vacondio2021grand}.
%
\added{These three concepts, tightly linked to each other, determine the accuracy of any numerical method. Consistency has been reasonably treated in the SPH literature
\cite{Quinlan_06}, even in the presence of boundaries; see, for instance, \cite{Maciaetal_PTP_2012, macia-etal_2011, merino2013consistency}. Convergence is determined by both consistency and stability. In SPH, stability issues are probably the most obvious, with most ``plain'' implementations quickly developing instabilities.}
Much effort has been devoted to alleviating these instabilities. The different techniques developed to this aim include artificial viscosity \cite{mon1992}, X-SPH \cite{monaghan2000sph}, Shepard filtering \cite{Quinlan_06}, zero-energy modes correction \cite{vignjevic2000}, Riemann solvers \cite{vila1999}, $\delta$-SPH \cite{antuono2021delta}, particle shifting \cite{Xu20096703}, incompressible-SPH \cite{lee2008}, and ALE \cite{oger2016sph}.
Except for the incompressible-SPH approaches, which deeply change the methodology, all the methods above share the common factor of adding ``artificial'' diffusion in order to produce smoother pressure fields.
However, a different approach, based on \added{analyzing the impact of time integration in} energy conservation, has recently demonstrated that the model can be unconditionally stable provided that an appropriate time integration scheme is selected \cite{cercospita_2023_energy}.
No boundaries were considered in such analysis.

Despite the connection between the two issues, very few publications target both BC and stability simultaneously.
For instance, \added{Mayrhofer \text{et al}} \cite{mayrhofer_etal_na_2014} address the skew-adjointness of the operators in BI, even proposing an alternative formulation which was left untested.
The energy conservation of FE schemes was also analyzed in Ref. \cite{CercosPita2016b}, indeed exposing some flaws.

This work presents a kind of boundary condition which is able to extend the unconditional stability of SPH (\added{shown in our previous article \cite{cercospita_2023_energy}}) to problems with wall BC.
\added{The proposal is formulated within the BI framework, but in order to avoid confusion, the term ``BI'' will be reserved for the previous method, as described in this reference.}

\added{Section \ref{s:gov_equations} presents the governing equations. The treatment of boundaries in SPH is introduced in Section \ref{s:sph} and the energy conservation in solid walls is discussed in Section \ref{s:energy}. Section \ref{s:applications} presents a collection of application cases where the solution proposed herein is tested successfully. These cases include a normal impact (Section \ref{ss:applications:normal_impact}), an adiabatic expansion (Section \ref{ss:applications:adiabatic_expansion}), a dam break test (Section \ref{ss:applications:dam_break}) and the water landing of a spacecraft 
(Section \ref{ss:applications:normal_impact}). Finally, Section \ref{s:conclusions} collects conclusions and ideas for future work.}


\section{Governing equations}
\label{s:gov_equations}
Although a similar analysis can be carried out for the incompressible case, as recently demonstrated in \cite{17th_SPHERIC_Merino_Violeau}, herein the focus is set on compressible flows, either weakly-compressible or fully compressible.
Hence, the governing equation for the evolution of density is the conservation equation:
\begin{equation}
\label{eq:gov_equations:mass_cons}
\SPH{ \frac{\D \rho }{ \D t} } _i(t) = -\rho_i(t) \SPH{\Div{\bs{u}}}_i(t)
\end{equation}
where $\SPH{\cdots}$ denotes \added{discrete} SPH operators and, abusing the notation, any magnitude resulting from the application of the SPH methodology.
In the above equation $\rho_i$ is the density of the $i$-th particle, and $\bs{u}_i$, its velocity.

For the sake of simplicity, non-viscous flows without volumetric forces are considered, such that the evolution of the velocity field is a discrete version of the Euler momentum equation,
\begin{equation}
\label{eq:gov_equations:mom_cons}
	\SPH{ \frac{\D \bs{u} }{\D t} }_i(t) =
		- \frac{\SPH{\Grad{p}}_i(t)}{\rho_i(t)},
\end{equation}
where $p_i$ is the pressure \added{of the $i$-th particle}.

In WC-SPH, these equations are closed by an equation of state (EOS) relating pressure and density:
\begin{equation}
\label{eq:gov_equations:eos}
 p_i(t) = p_0 + c^2_0 \, (\rho_i(t) - \rho_0),
\end{equation}
where $c_0$ is an artificial speed of sound, whose value should be high enough that density variations are small: \added{a departure of the density from its reference value $\rho_0$ causes a variation of the pressure from its reference value $p_0$, and the linear relationship is controlled by $c^2_0$. A high speed of sound causes a steeper variation of pressure.}

\section{Boundaries in SPH}
\label{s:sph}

The entire SPH formalism may be built upon a convolution: for
any arbitrary function $f(\bs{x})$ on $\mathbb{R}^{d}$, its \added{integral} SPH smoothed-out version is 
\begin{equation}
\label{eq:sph-integral}
\SPHint{f}(\bs{x}) = \int_{\mathbb{R}^{d}} f(\bs{y}) W(\bs{y} - \bs{x}) \D \bs{y},
\end{equation}
where $W(\bs{x})$ is a given kernel function with compact support, $\mathcal{B}(\bs{x})$. 
%
\added{The notation $\SPHint{\cdot}$ is used herein to refer the integral version of the SPH operators, as opposed to their discrete forms.}

Consider splitting the whole space, $\mathbb{R}^{d}$, so that we keep the computational domain, $\Omega$, and the \added{interior of its} complementary, which we will denote by ``extended domain'', $\bar{\Omega}$.
Note that both sets are open, that is, the boundary between them $\partial \Omega$ is not included in either $\Omega$ or $\bar{\Omega}$.
%
Thus the convolution in \eqref{eq:sph-integral} can be re-written as
\begin{equation}
\SPHint{f}(\bs{x}) =
\int_{\Omega \cup \bar{\Omega}} 
                    f(\bs{y}) W(\bs{y} - \bs{x}) \D \bs{y} =
\int_{\Omega}       f(\bs{y}) W(\bs{y} - \bs{x}) \D \bs{y} +
\int_{\bar{\Omega}} f(\bs{y}) W(\bs{y} - \bs{x}) \D \bs{y}.
\end{equation}

This convolution is not really useful for functions but gains significance when applied to differential operators. For example, for the gradient, \added{the expression yields}
\begin{equation}
\label{eq:sph:general_prev}
\SPHint{\Grad{f}}(\bs{x}) =
\int_{\Omega}       \Grad{f}(\bs{y}) W(\bs{y} - \bs{x}) \D \bs{y} +
\int_{\bar{\Omega}} \Grad{f}(\bs{y}) W(\bs{y} - \bs{x}) \D \bs{y}
\end{equation}
These terms can be rearranged applying the divergence theorem:
\begin{align}
\begin{split}\label{eq:sph:general}
\SPHint{\Grad{f}}(\bs{x}) = 
& - \int_{\Omega} f(\bs{y}) \Grad{W}(\bs{y} - \bs{x}) \D \bs{y} + \int_{\p \Omega} f(\bs{y}) \bs{n}(\bs{y}) W(\bs{y} - \bs{x}) \D \bs{y} \\
& - \int_{\bar{\Omega}} f(\bs{y}) \Grad{W}(\bs{y} - \bs{x}) \D \bs{y} + \int_{\p \bar{\Omega}} f(\bs{y}) \bs{n}(\bs{y}) W(\bs{y} - \bs{x}) \D \bs{y},
\end{split}
\end{align}
where $\bs{n}$ is the outward-pointing unit normal at each point on $\partial \Omega$.

Gauss's theorem applies only if the function $f$ is continuous in each subdomain, $\Omega$ and $\bar{\Omega}$, separately.
This permits defining a piecewise function that is different on both subdomains.
If the function $f$ is extended to the complementary subdomain, $\bar{\Omega}$, in such a way that it is continuous everywhere, then the boundary integrals are canceled out.
This is revisited below --- for the time being, the general expression \eqref{eq:sph:general} will be kept.

It is also worth noting that Eq. \eqref{eq:sph:general} consistently vanishes if the function $f$ is constant.
Moreover, the integrals in each subdomain cancel their boundary integral if $f$ is constant in such subdomain (i.e., each line of \eqref{eq:sph:general} vanishes independently).
This allows for the creation of symmetric and antisymmetric versions of the differential operators, which is in fact a common practice in order to achieve intrinsic momentum and energy conservation \cite{cercospita_2023_energy, monaghan_arfm_2012}.

In any case, the last three terms on the right-hand side of Eq. \eqref{eq:sph:general} conform the boundary terms in SPH, for which several methods have been proposed.
However, before addressing the different models for these terms and their role in energy balance, it is convenient to define the specific SPH operators in the absence of boundaries.

\subsection{SPH operators without boundaries}
\label{ss:sph:no_boundaries}

Far from the boundary (i.e.  farther than the kernel support distance), the \added{integral} SPH operators
may be written as
\begin{align}
\SPHint{\Grad{p}}^{\Omega}(\bs{x}) = 
- \int_{\Omega} \left(p(\bs{y}) + p(\bs{x})\right) \Grad{W}(\bs{y} - \bs{x}) \D \bs{y},
\\
\SPHint{\Div{\bs{u}}}^{\Omega}(\bs{x}) = 
- \int_{\Omega} \left(\bs{u}(\bs{y}) - \bs{u}(\bs{x})\right) \cdot \Grad{W}(\bs{y} - \bs{x}) \D \bs{y}.
\end{align}

Those operators can be discretized as
\begin{align}
\label{eq:sph:no_boundaries:gradp}
\SPH{\Grad{p}}^{\Omega}_i = 
- \sum_{j \in \Omega} \left(p_j + p_i\right) \Grad{W}_{ij} \frac{m_j}{\rho_j},
\\
\label{eq:sph:no_boundaries:divu}
\SPH{\Div{\bs{u}}}^{\Omega}_i = 
- \sum_{j \in \Omega} \left(\bs{u}_j - \bs{u}_i\right) \cdot \Grad{W}_{ij} \frac{m_j}{\rho_j}.
\end{align}
with $\Grad{W}_{ij}$ denoting $\Grad{W}(\bs{r}_j - \bs{r}_i)$.
In the expression above, $m_j$ is the mass of an arbitrary $j$-th particle, which is considered to be constant in time in this article.

Note that these operators are skew-adjoint: while in the continuum
\begin{equation}
	 \int_{\mathbb{R}^d} \left( \nabla p(\bs{y}) \right)  \cdot \bs{u}(\bs{y})  \D \bs{y}  =
-\int_{\mathbb{R}^d} \left( \nabla \cdot \bs{u}(\bs{y}) \right) p(\bs{y})   \D \bs{y} ,
\end{equation}
the discrete SPH operators of Eqs.~(\ref{eq:sph:no_boundaries:gradp},\ref{eq:sph:no_boundaries:divu}), satisfy \added{(far enough from the boundary) the corresponding discrete identity:}
\begin{equation}\label{eq:sph:no_boundaries:skewadjoint}
\sum_{i \in \Omega} 
\SPH{\Grad{p}}^\Omega_i \cdot \bs{u}_i  \frac{m_i}{\rho_i}
 =
-\sum_{i \in \Omega} 
\SPH{\Div{\bs{u}}}^\Omega_i p_i  \frac{m_i}{\rho_i} .
\end{equation}
\added{In absence of boundaries,} this property leads to accurate energy conservation if an appropriate time integration scheme is chosen \cite{cercospita_2023_energy}.

The boundary model consists of adding terms to Eqs.~(\ref{eq:sph:no_boundaries:gradp},\ref{eq:sph:no_boundaries:divu}) to approximate the missing last three terms of Eq. \eqref{eq:sph:general}. The most important methods are discussed next, continuing the discussion sketched in the Introduction.

\subsection{Boundary models}
\label{ss:sph:boundaries}
The SPH research community has proposed a number of ways to address boundary terms.
They may be grouped into three main categories.

\subsubsection{Boundary Forces (BF)}
\label{sss:sph:boundaries:bforces}

This first group would be the most na\"ive and with the weakest underlying formalism.
The most popular variants can be found on Refs. \cite{cercospita_thesis_2016, Monaghan_kajtar_cpc_2009, crespo2007b}.
A set of markers is placed either along the computational boundary $\partial \Omega$ or the complementary domain $\bar{\Omega}$, such that particles interacting with them experience a boundary force:
\begin{equation}\label{eq:sph:boundaries:bforces}
\SPH{\Grad{p}}_i = 
- \sum_{j \in \Omega} \left(p_j + p_i\right) \Grad{W}_{ij} \frac{m_j}{\rho_j} + \sum_{j \in \partial \Omega \cup \bar{\Omega}} \bs{f}_j,
\end{equation}
where the second term is the force, with $\bs{f}_j$ tuned in order to achieve the desired effect.
These models usually show poor performance, so they have historically evolved into models that fall into the other categories.
As an example, the Dynamic Boundary Conditions (DBC) discussed in Ref. \cite{crespo2007b} evolved to a fluid extension called modified DBC (mDBC) in Ref. \cite{english_2022_mDBC}.

\subsubsection{Fluid Extensions (FE) }
\label{sss:sph:boundaries:gp}

This category has sometimes been split into two subcategories: Dummy Particles and Ghost Particles.
However, the reasoning behind both is actually the same: populating the complementary domain, $\bar{\Omega}$, with ``extra'' particles and defining extended fields for them. This allows to write the SPH operators as
\begin{align}
\label{eq:sph:boundaries:gp:gradp}
\SPH{\Grad{p}}_i 
&= 
- \sum_{j \in \Omega} \left(p_j + p_i\right) \Grad{W}_{ij} \frac{m_j}{\rho_j}
- \sum_{j \in \bar{\Omega}} \left(p_j + p_i\right) \Grad{W}_{ij} \frac{m_j}{\rho_j},
\\
\label{eq:sph:boundaries:gp:divu}
\SPH{\Div{\bs{u}}}_i 
&= 
- \sum_{j \in \Omega} \left(\bs{u}_j - \bs{u}_i\right) \cdot \Grad{W}_{ij} \frac{m_j}{\rho_j}
- \sum_{j \in \bar{\Omega}} \left(\bs{u}_j - \bs{u}_i\right) \cdot \Grad{W}_{ij} \frac{m_j}{\rho_j}.
\end{align}
\added{Note that these operators are different from those defined by Eqs. \eqref{eq:sph:no_boundaries:gradp} and \eqref{eq:sph:no_boundaries:divu} in the fact that those considered only particles within $\Omega$, while these include the particles in $\bar{\Omega}$. The two definitions are the same far away from the boundaries or in the absence of them, as explained in Section \ref{ss:sph:no_boundaries}. Also, note that this} is a rather lax notation, since for extended particles, with $j \in \bar{\Omega}$, their actual position and field values may depend upon $i$, as will indeed happen later on.

As aforementioned, to drop the boundary integrals of Eq. (\ref{eq:sph:general}),
$\rho_j$, $p_j$, and $\bs{u}_j$ should be defined on $\bar{\Omega}$ so that they are continuous on the border $\partial \Omega$.
That might actually be the case or not.
However, this eventuality has traditionally been circumvented by just asking for a consistency property to be fulfilled in the limit when the SPH kernel approaches a Dirac $\delta$ function. This may be formally defined through the characteristic standard deviation of the kernel:
\begin{equation}
	\sigma^2 = 
 \int_{\mathbb{R}^{d}} |\bs{x}|^2 W(\bs{x}) \D \bs{x},
\end{equation}
with the consistency requirement
\begin{equation}
\label{eq:sph:boundaries:gp:gradp_consistency}
\lim_{\sigma \rightarrow 0} \SPH{\Grad{p}}_i = \Grad{p}(\bs{x}_i)\qquad
\lim_{\sigma \rightarrow 0} \SPH{\Div{\bs{u}}}_i = \Div{\bs{u}}(\bs{x}_i) 
\end{equation}
for all $\bs{x}_i \in \partial \Omega$.

Several examples of definitions based on those consistency requirements can be found on the literature,
for instance in Refs. \cite{8th_SPHERIC_Merino_etal, MaciaetalPTP, Marrone2011, antuono2023clone}.
It is also worth mentioning that the impact of fluid extensions upon energy balance was explored in Ref. \cite{CercosPita2016b}, and briefly revisited in Ref. \cite{cercospita_thesis_2016}.
However, such energy balance is addressed here once again, following the guidelines on energy balance discussed on Ref. \cite{cercospita_2023_energy}

\subsubsection{Boundary Integrals (BI)}
\label{sss:sph:boundaries:bi}

Boundary Integrals (BI) are built in a somewhat artificial way, by accepting the following approximation to differential operators in the outside subdomain:
\begin{equation}
\Grad{f}(\bs{y}) W(\bs{y} - \bs{x}) 
\simeq \SPH{\Grad{f}}(\bs{x}) \, W(\bs{y} - \bs{x}) 
\qquad \forall \bs{x} \in \Omega ,  \bs{y} \in \bar{\Omega} .
\end{equation}
%
The eventual consequences of such approximation are further discussed in \ref{s:bi_approx}.
Under such approximation, the general gradient of Eq.~\eqref{eq:sph:general_prev} may be computed as
\begin{equation}
\label{eq:sph:boundaries:bi:general_prev}
\SPH{\Grad{f}}(\bs{x}) = \int_{\Omega} \Grad{f}(\bs{y}) W(\bs{y} - \bs{x}) \D \bs{y} + \SPH{\Grad{f}}(\bs{x}) \int_{\bar{\Omega}} W(\bs{y} - \bs{x}) \D \bs{y}.
\end{equation}
The Shepard renormalization factor may now be invoked,
\begin{equation}
	\label{eq:sph:shepard}
	\gamma(\bs{x}) = \int_{\Omega} W(\bs{y} - \bs{x}) \D \bs{y},
\end{equation}
which equals $1$ away from the boundaries, but will diminish close to them, as part of the kernel support will lie outside the fluid domain. That part of the support is precisely the integral $\int_{\bar{\Omega}} W(\bs{y} - \bs{x}) \D \bs{y}$, which must therefore equal $ 1 - \gamma(\bs{x})$, hence Eq.~\eqref{eq:sph:boundaries:bi:general_prev} can be written as
\begin{equation}
\SPH{\Grad{f}}(\bs{x}) = \frac{1}{\gamma(\bs{x})} \int_{\Omega} \Grad{f}(\bs{y}) W(\bs{y}) .
\end{equation}
This can be rearranged applying Gauss' theorem,
\begin{equation}
\label{eq:sph:boundaries:bi:general}
\SPH{\Grad{f}}(\bs{x}) = 
- \frac{1}{\gamma(\bs{x})} \int_{\Omega}  f(\bs{y}) \Grad{W}(\bs{y} - \bs{x}) \D \bs{y} +
  \frac{1}{\gamma(\bs{x})} \int_{\partial \Omega} f(\bs{y}) \bs{n}(\bs{y}) W(\bs{y} - \bs{x}) \D \bs{y}.
\end{equation}

Two main approaches have been proposed in order to discretize the problem.
In Ref. \cite{ferrand_etal_2012}, the boundary was discretized in a set of lines in 2D (3D formulations were not provided), and the integral along the boundary, $\partial \Omega$, was solved in a semi-analytical way.
In contrast, Refs. \cite{Maciaetal_PTP_2012, cercospita_thesis_2016} employed a purely discrete approach, consisting of populating the boundary with a set of boundary particles, each representing the area of its boundary portion.

Whereas the former can result in slightly more accurate results, its added complexity makes it less attractive for a wide spectrum of practical applications.
Herein we are considering the latter, which after discretizing results in the following operators:
\begin{align}
\label{eq:sph:boundaries:bi:gradp}
\SPH{\Grad{p}}_i = 
- \frac{1}{\gamma_i} \sum_{j \in \Omega} \left(p_j + p_i\right) \Grad{W}_{ij} \frac{m_j}{\rho_j}
+ \frac{1}{\gamma_i} \sum_{j \in \partial \Omega} \left(p_j + p_i\right) \bs{n}_j W_{ij} s_j,
\\
\label{eq:sph:boundaries:bi:divu}
\SPH{\Div{\bs{u}}}_i = 
- \frac{1}{\gamma_i} \sum_{j \in \Omega} \left(\bs{u}_j - \bs{u}_i\right) \cdot \Grad{W}_{ij} \frac{m_j}{\rho_j}
+ \frac{1}{\gamma_i} \sum_{j \in \partial \Omega} \left(\bs{u}_j - \bs{u}_i\right) \cdot \bs{n}_j W_{ij} s_j,
\end{align}
where $\bs{n}_j$ and $s_j$ are the outward-pointing normal and area of an arbitrary $j$-th boundary element.
\section{Energy conservation at solid walls}
\label{s:energy}

If an adequate time integration scheme is selected, energy conservation reduces to power conservation \cite{cercospita_2023_energy}.
More specifically, for the governing equations of Section \ref{s:gov_equations}, it can be asserted that
\begin{equation}
	\label{total_power}
\SPH{P_\text{k}} + \SPH{P_\text{c}} = -\SPH{P}^{\partial \Omega},
\end{equation}
where $\SPH{P}^{\partial \Omega}$ is the mechanical energy transferred by the fluid to the solid walls, and the power associated to the kinetic energy and compressibility are defined as:
\begin{align}
\SPH{P_\text{k}} :=
\sum_{i \in \Omega} m_i \bs{u}_i \cdot \SPH{\frac{\D \bs{u}}{\D t}}_i = -\sum_{i \in \Omega} \frac{m_i}{\rho_i} \bs{u}_i \cdot \SPH{\Grad{p}}_i,
\\
\SPH{P_\text{c}} :=
\sum_{i \in \Omega} m_i \frac{p_i}{\rho_i^2} \SPH{\frac{\D \rho}{\D t}}_i =
-\sum_{i \in \Omega}  \frac{m_i}{\rho_i} p_i \SPH{\Div{\bs{u}}}_i .
\end{align}

The kinetic and compressible powers can be conveniently split into the powers due to fluid-fluid interactions and those due to fluid-boundary interactions, i.e.
\begin{align}
\SPH{P_\text{k}} := - \sum_{i \in \Omega} \frac{m_i}{\rho_i} \bs{u}_i \cdot \SPH{\Grad{p}}^{\Omega}_i - \sum_{i \in \Omega} \frac{m_i}{\rho_i} \bs{u}_i \cdot \SPH{\Grad{p}}^{\partial \Omega}_i,
\\
\SPH{P_\text{c}} := - \sum_{i \in \Omega}  \frac{m_i}{\rho_i} p_i \SPH{\Div{\bs{u}}}^{\Omega}_i - \sum_{i \in \Omega}  \frac{m_i}{\rho_i} p_i \SPH{\Div{\bs{u}}}^{\partial \Omega}_i,
\end{align}
with $\SPH{\ldots}^{\partial \Omega} = \SPH{\ldots} - \SPH{\ldots}^{\Omega}$.

When these two powers are added in Eq.~\eqref{total_power}, the first terms on each of the right hand sides cancel each other out, thanks to Eq.~\eqref{eq:sph:no_boundaries:skewadjoint}. Hence, intrinsic energy conservation in interactions with solid walls reduces to the definition of suitable SPH operators which result in meaningful $\SPH{P}^{\partial \Omega}$ terms.
For the sake of simplicity, the problem is now divided into two cases: fixed walls,
when the fluid-wall mechanical energy transfer vanishes, and moving walls, where it does not.
%

\subsection{Fixed walls}
\label{ss:energy:homogeneous}
Stationary solid walls have played a central role in the development of CFD in general, and of SPH in particular.
In our context, fixed walls are characterized by $\SPH{P}^{\partial \Omega} = 0$.

The BI formulation described on Section \ref{sss:sph:boundaries:bi} results in
boundary operators that may be written as the whole SPH expression minus a bulk term (as if the boundary did not exist):
\begin{align}
\SPH{\Grad{p}}_i^{\partial \Omega} &=
\SPH{\Grad{p}}_i -
\left(
- \sum_{j \in \Omega} \left(p_j + p_i\right) \Grad{W}_{ij} \frac{m_j}{\rho_j}
\right)
\\
\SPH{\Div{\bs{u}}}_i^{\partial \Omega} & = 
\SPH{\Div{\bs{u}}}_i-
\left(
- \sum_{j \in \Omega} \left(\bs{u}_j - \bs{u}_i\right) \cdot \Grad{W}_{ij} \frac{m_j}{\rho_j}
\right) .
\end{align}
These may be rewritten as
\begin{align}
\SPH{\Grad{p}}_i^{\partial \Omega} &= 
- \frac{1 - \gamma_i}{\gamma_i} \sum_{j \in \Omega} \left(p_j + p_i\right) \Grad{W}_{ij} \frac{m_j}{\rho_j}
+ \frac{1}{\gamma_i} \sum_{j \in \partial \Omega} \left(p_j + p_i\right) \bs{n}_j W_{ij} s_j,
\\
\SPH{\Div{\bs{u}}}_i^{\partial \Omega} &= 
- \frac{1 - \gamma_i}{\gamma_i} \sum_{j \in \Omega} \left(\bs{u}_j - \bs{u}_i\right) \cdot \Grad{W}_{ij} \frac{m_j}{\rho_j}
+ \frac{1}{\gamma_i} \sum_{j \in \partial \Omega} \left(\bs{u}_j - \bs{u}_i\right) \cdot \bs{n}_j W_{ij} s_j.
\end{align}  
Unsurprisingly, the presence of the renormalization factor produces a term on each SPH boundary operator that depends on fluid-fluid interactions (first term of each equation).
Unfortunately, 
these operators are not skew-adjoint anymore.
%
%
It therefore seems impossible to define BI operators that intrinsically conserve energy.
Even if this can be achieved through an iterative process,
in this work this possibility will not be pursued, 
and SPH operators involving renormalization factors will no longer be considered.

In contrast the boundary SPH operators in the FE approach read
\begin{align}
\SPH{\Grad{p}}_i^{\partial \Omega} = 
- \sum_{j \in \bar{\Omega}} \left(p_j + p_i\right) \Grad{W}_{ij} \frac{m_j}{\rho_j},
\\
\SPH{\Div{\bs{u}}}_i^{\partial \Omega} = 
- \sum_{j \in \bar{\Omega}} \left(\bs{u}_j - \bs{u}_i\right) \cdot \Grad{W}_{ij} \frac{m_j}{\rho_j},
\end{align}
such that the mechanical energy given to the boundary has a rate
\begin{equation}
\label{eq:energy:homogeneous:gp}
\SPH{P}^{\partial \Omega} = -\sum_{i \in \Omega} \sum_{j \in \bar{\Omega}} \frac{m_i}{\rho_i} \frac{m_j}{\rho_j} \left(p_j \bs{u}_i + p_i \bs{u}_j \right) \cdot \Grad{W}_{ij}.
\end{equation}
which indeed can be canceled by choosing appropriate field extensions, $\bs{u}_j$ and $p_j$.

Indeed, this expression 
vanishes for infinite symmetry planes, as discussed in \ref{s:sym_plane}.
However, in order to ensure this rate vanishes for more general wall shapes, extended fields of the following form may be defined:
\begin{equation} \label{eq:vel_and_press_at_boundary}
\begin{cases}
\bs{u}_j = - k_i \bs{u}_i \\
p_j = k_i p_i,
\end{cases}
\end{equation}
with $k_i$ an arbitrary constant which can change from particle to particle, and in time.
This clearly resembles the so-called clone particles approach \cite{antuono2023clone}.
Notice that in our lax notation the fields in the extended particle $j$ actually depend on $i$:  each inside particle ``sees'' different fields in the extended particles.

From all the possible values that $k_i$ can take, two particular ones have a clear physical meaning:
\begin{itemize}
    \item $k_i = -1$: Free surface model. Informally, the velocity in the extended subdomain is the same as in the inside subdomain. This mimics a phase coexistence between our physical fluid and a fictitious one outside. The pressure is, informally, an odd function across the boundary, and therefore should be zero on it.
    \item $k_i = 1$: Local symmetric extension. Also informally, the velocity at the boundary is zero, whereas the pressure is an even function. This mimics the usual no-slip boundary condition for the velocity and zero normal gradient for the pressure.
\end{itemize}

The latter choice is clearly the sensible one to model fixed walls and will be the only one considered.
It leads to a pair of boundary operators which grant energy conservation,
while retaining 0-th order consistency:
\begin{align}
\label{eq:energy:homogeneous:gradp_gp}
\SPH{\Grad{p}}_i^{\partial \Omega} = 
- 2 p_i \sum_{j \in \bar{\Omega}} \Grad{W}_{ij} \frac{m_j}{\rho_j},
\\
\label{eq:energy:homogeneous:divu_gp}
\SPH{\Div{\bs{u}}}_i^{\partial \Omega} = 
2 \bs{u}_i \cdot \sum_{j \in \bar{\Omega}} \Grad{W}_{ij} \frac{m_j}{\rho_j}.
\end{align}

Even though we have deviated from the traditional BI formulation, a direct application of Gauss's theorem allows us to write these expressions in a BI fashion.
The rate of Eq.~\eqref{eq:energy:homogeneous:gp} becomes
\begin{equation}
\label{eq:energy:homogeneous:gp_bi}
\SPH{P}^{\partial \Omega} =
\sum_{i \in \Omega} \sum_{j \in \partial \Omega}
\frac{m_i}{\rho_i}
\left(p_j \bs{u}_i + p_i \bs{u}_j \right) \cdot 
\bs{n_j} W_{ij} s_j,
\end{equation}
which again vanishes when the conditions of Eq.~\eqref{eq:vel_and_press_at_boundary} are satisfied.
Taking the $k_i=1$ choice, we find
\begin{align}
\label{eq:energy:homogeneous:gradp_bi}
\SPH{\Grad{p}}_i^{\partial \Omega} = 
2 p_i \sum_{j \in \partial \Omega} \bs{n_j} W_{ij} s_j,
\\
\label{eq:energy:homogeneous:divu_bi}
\SPH{\Div{\bs{u}}}_i^{\partial \Omega} = 
- 2 \bs{u}_i \cdot \sum_{j \in \partial \Omega} \bs{n_j} W_{ij} s_j.
\end{align}
These expressions constitute the main contribution of this work,
as far as fixed walls are concerned.
\added{%
	It must be stressed that the resulting method clearly retains its flavor of a ``boundary integral'' technique	even if arguments from FE have been used in its derivation. Its possible implementation in FE is appealing, but it is not pursued here. We will also reserve the term ``BI'' for the previous formulation of the method, as in Eqs.~(\ref{eq:sph:boundaries:bi:gradp}, \ref{eq:sph:boundaries:bi:divu}).}

Note that conditions Eq.~\eqref{eq:vel_and_press_at_boundary} are sufficient, but less strict choices may be considered. For instance, the choice
%
\begin{equation}
	p_j =   \alpha_j \sum_{i \in \Omega}  \frac{m_i}{\rho_i} W_{ij}  p_i ,
\end{equation}
together with an equivalent expression for the velocity, is also sufficient to produce a vanishing wall energy rate. The sensible choice would now be $\alpha_j=2$, in order to connect with the force of Eq. \eqref{capital_j_def}. Unfortunately, the application of those alternative formulations would not yield operators compatible with FE, as a direct application of Gauss's theorem would no longer be feasible.

\added{%
0-th order consistency is not enough to enforce that the wall is not trespassed in some cases.
To this end, BF may be introduced so that in the numerical scheme the normal component of the velocities is reversed if needed.
In other words, the particles about to trespass the boundary can be subject to elastic forces
--- no mechanical work is made, since these forces do not change the velocity magnitude.
This approach has been explored, and integrated
within a conservative implicit scheme \cite{cercospita_2023_energy},
but results are not reported since the impact of this technique is quite
negligible. It may likely become important in cases where there is significant loss of particles across the walls.
}

%
%

\subsection{Moving walls}
\label{ss:energy:particular}
Although in this case  $\SPH{P}^{\partial \Omega} \neq 0$, the argument in the previous section can be actually extended to moving walls in a natural way.
With the \textit{a priori} knowledge of the result we can guess the following differential operators:
\begin{align}
\SPH{\Grad{p}}_i^{\partial \Omega} = 
\sum_{j \in \partial \Omega} \left(p_j + p_i\right) \bs{n}_j W_{ij} s_j,
\\
\label{eq:energy:particular:duvu_bi_abs}
\SPH{\Div{\bs{u}}}_i^{\partial \Omega} = 
\sum_{j \in \partial \Omega} \left(\bs{u}_j - \bs{u}_i\right) \cdot \bs{n}_j W_{ij} s_j,
\end{align}
where, again, the fields $p_j$ and $\bs{u}_j$ must be suitably defined. 

As discussed above, for a fixed wall, the mechanical energy transferred to it can only vanish when the mechanical energy exchange due to each pair interaction does.
The mechanical energy transferred to the boundary in the BI
framework is given by Eq.~\eqref{eq:energy:homogeneous:gp_bi}. In it,
we may identify the mechanical energy exchanged on each pair interaction,
\begin{equation}
\SPH{P}^{\partial \Omega}_{ij} := \frac{m_i}{\rho_i} s_j \left(p_j \bs{u}_i + p_i \bs{u}_j \right) \cdot \bs{n}_j W_{ij}.
\end{equation}

Assume a moving wall described by the velocity of each $j$-th area element, $\bs{U}_j$, and write 
\begin{equation}
p_j \bs{u}_i + p_i \bs{u}_j =
\underbrace{p_j \left( \bs{u}_i - \bs{U}_j \right) + p_i \left( \bs{u}_j - \bs{U}_j \right) }
+ \left( p_i + p_j \right) \bs{U}_j .
\end{equation}
The underbraced term should vanish for moving walls, given that $\bs{u}_i - \bs{U}_j$ and $\bs{u}_j - \bs{U}_j$ are just velocities relative to the moving wall. We are therefore left with
\begin{equation}
	\begin{cases}
		\bs{u}_j = -\left(\bs{u}_i - 2 \bs{U}_j\right) \\
		p_j = p_i,
	\end{cases}
\end{equation}
akin to Eq. \eqref{eq:vel_and_press_at_boundary} (here, we only consider the equivalent $k_i=1$ case, which is the appropriate choice for walls.) 

This leads to a mechanical energy transfer rate
\begin{equation}
	\SPH{P}^{\partial \Omega}_{ij} = 
2 \frac{m_i}{\rho_i} \left( s_j p_i \bs{n}_j \right)  \cdot \bs{U}_j   W_{ij},
\end{equation}
a natural result:
\begin{equation*}
	\bs{F}_j = s_j \bs{n}_j 2 \sum_{i \in \Omega} p_i W_{ij} \frac{m_i}{\rho_i}
\end{equation*}
is the force due to pressure upon element $j$, and its scalar product with the element velocity, $\bs{F}_j\cdot \bs{U}_j  $, is the mechanical power at the element.

The resulting differential operators are
\begin{align}
\label{eq:energy:particular:gradp_bi}
\SPH{\Grad{p}}_i^{\partial \Omega} = 
2 p_i \sum_{j \in \partial \Omega} \bs{n_j} W_{ij} s_j,
\\
\label{eq:energy:particular:divu_bi}
\SPH{\Div{\bs{u}}}_i^{\partial \Omega} = 2
\sum_{j \in \partial \Omega}  \left(\bs{U}_j - \bs{u}_i \right) \cdot \bs{n_j} W_{ij} s_j.
\end{align}

Incidentally, it should be mentioned that a similar analysis can be drafted for FE instead of BI.
However, that way would lead to non-trivial questions, like the definition of the velocity $\bs{U}_j$, or the point of application of the forces.

%
%
%
%

\section{Numerical tests}
\label{s:applications}
The theoretical arguments above are tested here. First, a case with
a fixed wall is simulated: a normal impact of a fluid jet upon
a surface. The next case involves a moving piston. Then, a
canonical dam-break simulation is presented. \added{Finally, a simulation of spacecraft water landing is provided, as and advanced, fully 3D application, which includes the motion of the immersed body.}

\subsection{Normal impact}
\label{ss:applications:normal_impact}
\begin{figure}
	\centering
	\includegraphics[width=0.6\textwidth]{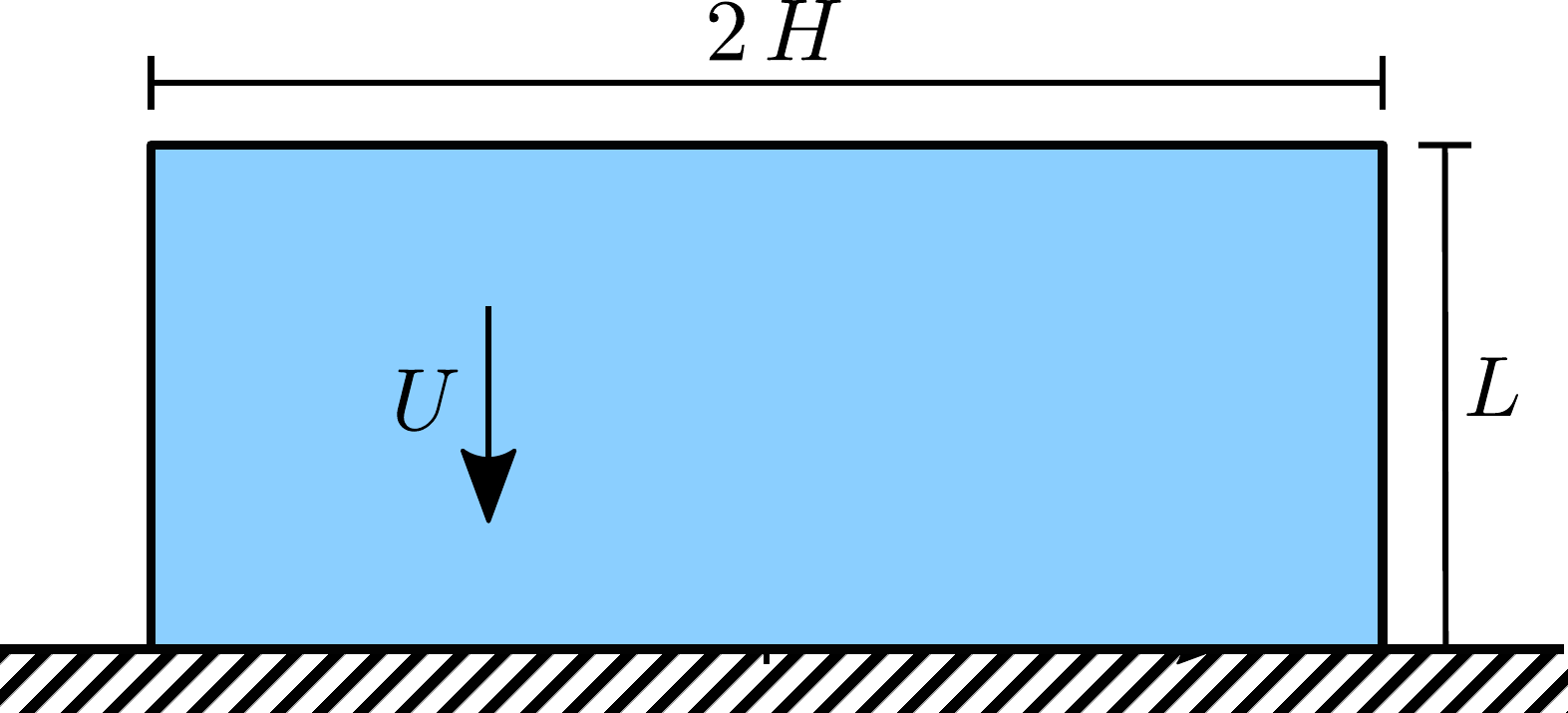}
	\caption{Schematic view of the fluid jet impact initial condition.}
	\label{fig:applications:normal_impact:scheme}
\end{figure}

A fluid jet of dimensions $2 H \times L$ impacts against a solid horizontal wall with a normal velocity $U$, as schematically depicted on Fig. \ref{fig:applications:normal_impact:scheme}.
The initial fluid density is equal to the reference one, $\rho(\boldsymbol{r}, t=0) = \rho_0 $.
No background pressure or volumetric forces are considered.
Thus, the whole phenomenon is governed by the Mach number
$\text{Ma} = U / c_0$.

This application was chosen because of its similarity to the impact of normal fluid jets, which is normally used to evaluate SPH energy conservation \cite{marrone2015, cercospita_2023_energy}.
For practical purposes, unity parameters are considered, i.e. $L = H = 1$, $\rho_0 = 1$, $U=1$, and $\text{Ma} = 0.01$.

The simulation is carried out with a semi-implicit midpoint time scheme, with $n_{\Delta t} = 15$ iterations per time step.
For the sake of optimization, the midpoint semi-implicit iterator is stopped when the energy residue,
\begin{equation}
\label{eq:applications:normal_impact:residue}
	R_{\Delta t}(t) = \sum_{i \in \Omega} \left\vert
		m_i \bs{u}_i(t) \cdot \left(
		\SPH{\frac{\D \bs{u}}{\D t}}_i(t)
	\right) \right\vert +
	\sum_{i \in \Omega} \left\vert
		\frac{m_i p_i(t_{n+1/2})}{\rho_i^2(t)} \left(\SPH{\frac{\D \rho}{\D t}}_i(t)
		\right) \right\vert
\end{equation}
satisfies $R_{\Delta t}(t_{n+1/2}) < 0.2 \times 10^{-2} \rho_0 L H U^2$.
No relaxation is considered in the semi-implicit iterator at the beginning of each time step, although a progressively increasing relaxation factor is introduced if the residue does not decrease at a geometric rate of $1/3$.

The fluid domain is discretized in a lattice of $N=80000$ particles with the same mass $m_i = 2 L H / N$, resulting in an initial inter-particle spacing of $\Delta x = 2 L / \sqrt{N}$.
A quintic Wendland kernel \cite{dehnen_aly_wendland_2012} is applied and its smoothing length $h$ is chosen so that $h / \Delta x = 4$.
The same inter-particle spacing $\Delta x$ is considered for the boundary elements.
The simulations run for a total time of three periods (the period is $ T = L \text{Ma} / U $), so there are no tensile instability issues.
The Courant number, $\text{Co} = h / c_0 \Delta t$, is set to $1/2$.

\begin{figure}[!ht]
	\centering
	\includegraphics[width=0.98\textwidth]{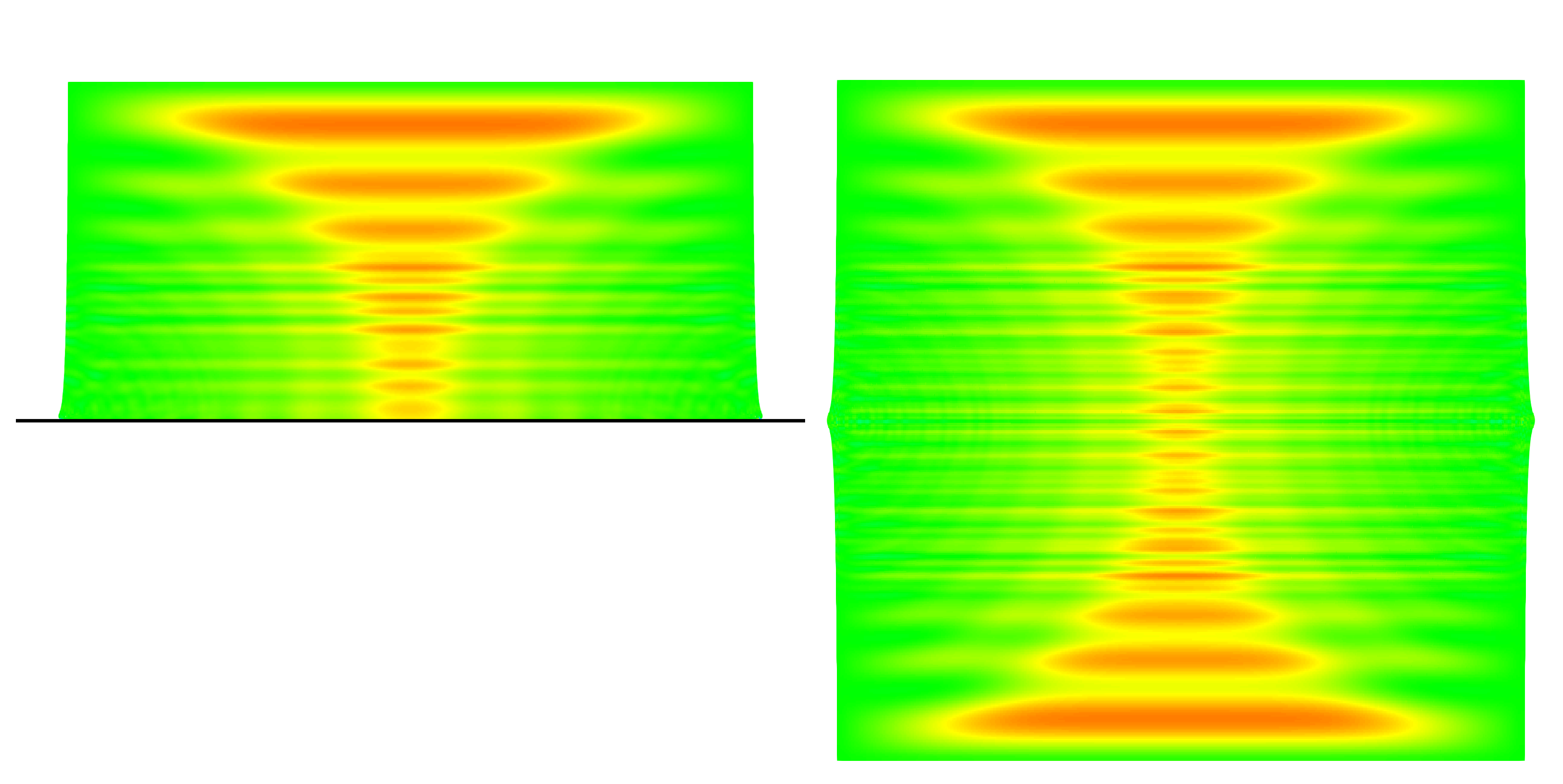}\\
    \input{normal_impact_snapshot_colorbar.pgf}
	\caption{Pressure field snapshot at $t = T / 4$. Left: Normal fluid jet impact against a solid wall. Right: Normal fluid jet impact \cite{cercospita_2023_energy}.}
	\label{fig:applications:normal_impact:snapshot}
\end{figure}

\begin{figure}[!ht]
	\centering
    \input{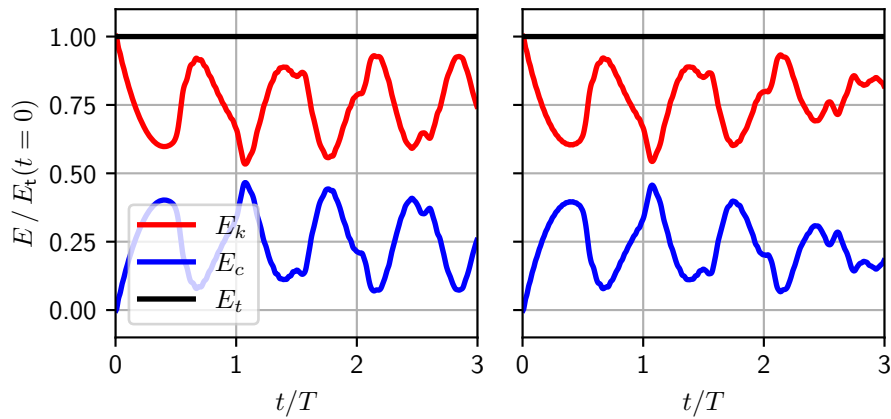}
	\caption{Energy profile. Left: Normal fluid jet impact against a solid wall. Right: Normal fluid jet impact \cite{cercospita_2023_energy}. Red: kinetic energy, blue: compression energy, black: total energy.}
	\label{fig:applications:normal_impact:energy}
\end{figure}

In Fig. \ref{fig:applications:normal_impact:snapshot} a snapshot of the solid wall simulation is shown and compared against the fluid jet impact.
In Fig. \ref{fig:applications:normal_impact:energy} the energy evolution for both simulations is depicted. 
As can be appreciated, replacing the bottom fluid by a solid wall consistently has a noticeable effect on the energy profile.
However, the formulation described above conserves energy by preventing any net transfer of energy between the fluid and the fixed wall, thereby ensuring unconditional stability.

%

%
\subsection{Adiabatic piston}
\label{ss:applications:adiabatic_expansion}
\begin{figure}[!ht]
	\centering
	\includegraphics[width=0.6\textwidth]{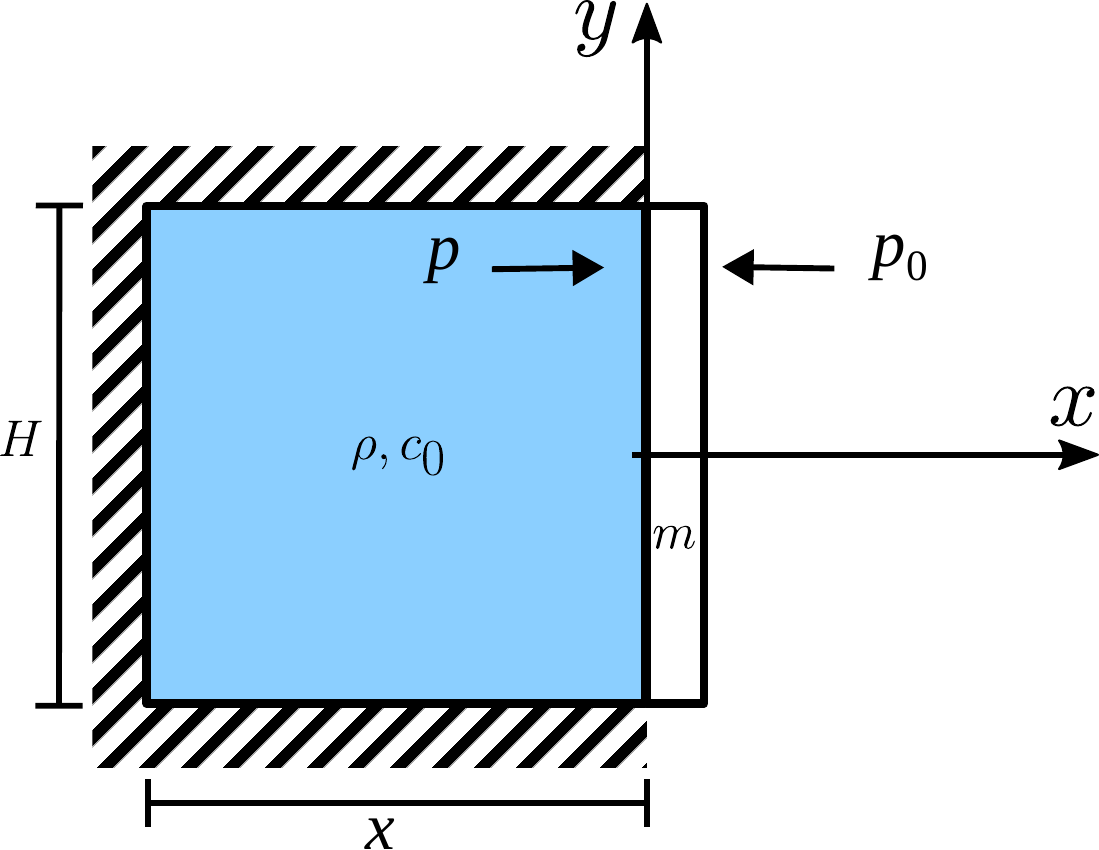}
	\caption{Schematic view of the adiabatic expansion.}
	\label{fig:applications:adiabatic_expansion:scheme}
\end{figure}

A second practical application consists in the simulation of R\"uchardt's experiment \cite{zemansky1998}.
A compressible fluid with density $\rho_0$ and speed of sound $c_0$ is enclosed in a rectangular tank of dimensions $x \times H$.
While three of the walls are fixed, a fourth wall consists of a piston with mass $m$ at position $x$, upon which a background pressure $p_0$ acts against the fluid.
The fluid pressure, $p$, balances the ambient pressure $p_0$ for a certain horizontal position $x_0$.
A schematic view of the simulation setup is shown in Fig. \ref{fig:applications:adiabatic_expansion:scheme}.
At the beginning of the simulation the piston is displaced a distance $A$ away from $x_0$, causing it to start oscillating around the equilibrium position.
The motion is described by
\begin{equation}
\label{eq:applications:adiabatic_expansion:EOM}
\frac{d^2 x}{dt^2} = \omega^2 x_0 \left( \frac{x_0}{x} -1 \right ),
\end{equation}
where
\begin{equation}
\label{eq:applications:adiabatic_expansion:omega}
 \omega^2 = \left( \frac{c_0}{x_0} \right)^2 \frac{M}{m},
\end{equation}
with $M$ the total mass of the fluid.
A typical initial condition is $x(0) = x_0 - A$ and $v(0)=0$.

This is a non-linear oscillator, but for small oscillations ($A \ll x_0$) the movement is harmonic, with an angular frequency $\omega$.
The resulting Mach number is, for small oscillations, given by
\begin{equation}
	\text{Ma} = \frac{A \omega}{c_0} = \frac{A}{x_0} \sqrt{\frac{M}{m}}.
\end{equation}

A second Mach number can be defined upon the initial compression ratio, $\rho/\rho_0$:
\begin{equation}
	\text{Ma}^{*} = \sqrt{\frac{\rho}{\rho_0} - 1} = \sqrt{\frac{A}{x_0 - A}}.
\end{equation}
For $\text{Ma} , \text{Ma}^{*}  \ll 1$, the fluid pressure is isotropic and its kinetic energy becomes negligible.

In the simulations, a square tank is considered, with $x_0 = H = 1$, filled with a fluid with a reference density of $\rho_0 = 1$, and a speed of sound of $c_0 = 1$.
Consequently, the mass of fluid enclosed in the tank is $M = \rho_0 / (x_0 H) = 1$, and the equilibrium position is $x = x_0$.
The simulation is carried out with Mach numbers $\text{Ma}^{*} = 0.1$ and $\text{Ma} = 4.5\times 10^{-3}$, in the harmonic regime, with an anisotropic fluid pressure profile and a non-negligible kinetic energy.
%
%
%
%
%
This sets $\omega = 0.45$ and a period of $T = 2\pi / \omega \approx 14$.

%
%
A background pressure equal to $3/2 c_0 (\rho(0) - \rho_0)$ is chosen to grant positive pressure values throughout the simulation.

The simulation is carried out with a semi-implicit midpoint time scheme, with $n_{\Delta t} = 4$ iterations per time step, without relaxation,
while he resulting energy residue of Eq. (\ref{eq:applications:normal_impact:residue}) is kept quite low, $R_{\Delta t}(t_{n+1/2}) < 1.0 \times 10^{-7} p_0 H x(t=0)$.
The wall motion is computed considering the same time integration scheme.

The fluid domain is discretized in a lattice of $N=400\times 400$ particles with the same mass $m_i = x_0 H / N$, resulting in an initial inter-particle spacing of $\Delta x = x_0 / \sqrt{N}$.
A quintic Wendland kernel \cite{dehnen_aly_wendland_2012} is applied and its smoothing length $h$ is chosen so that $h / \Delta x = 4$.
The same inter-particle spacing, $\Delta x$ is considered for the boundary elements.
However, while the left and right walls have a constant inter-boundary element distance, the distance between the boundary elements at the top and bottom walls is modified according to the motion of the piston, so a constant number of boundary elements is kept along the full simulation.
Up to three periods are simulated with a Courant number $\text{Co} = 2.5 \times 10^{-2}$ .

\begin{figure}[!ht]
	\centering
	\includegraphics[width=0.9\textwidth]{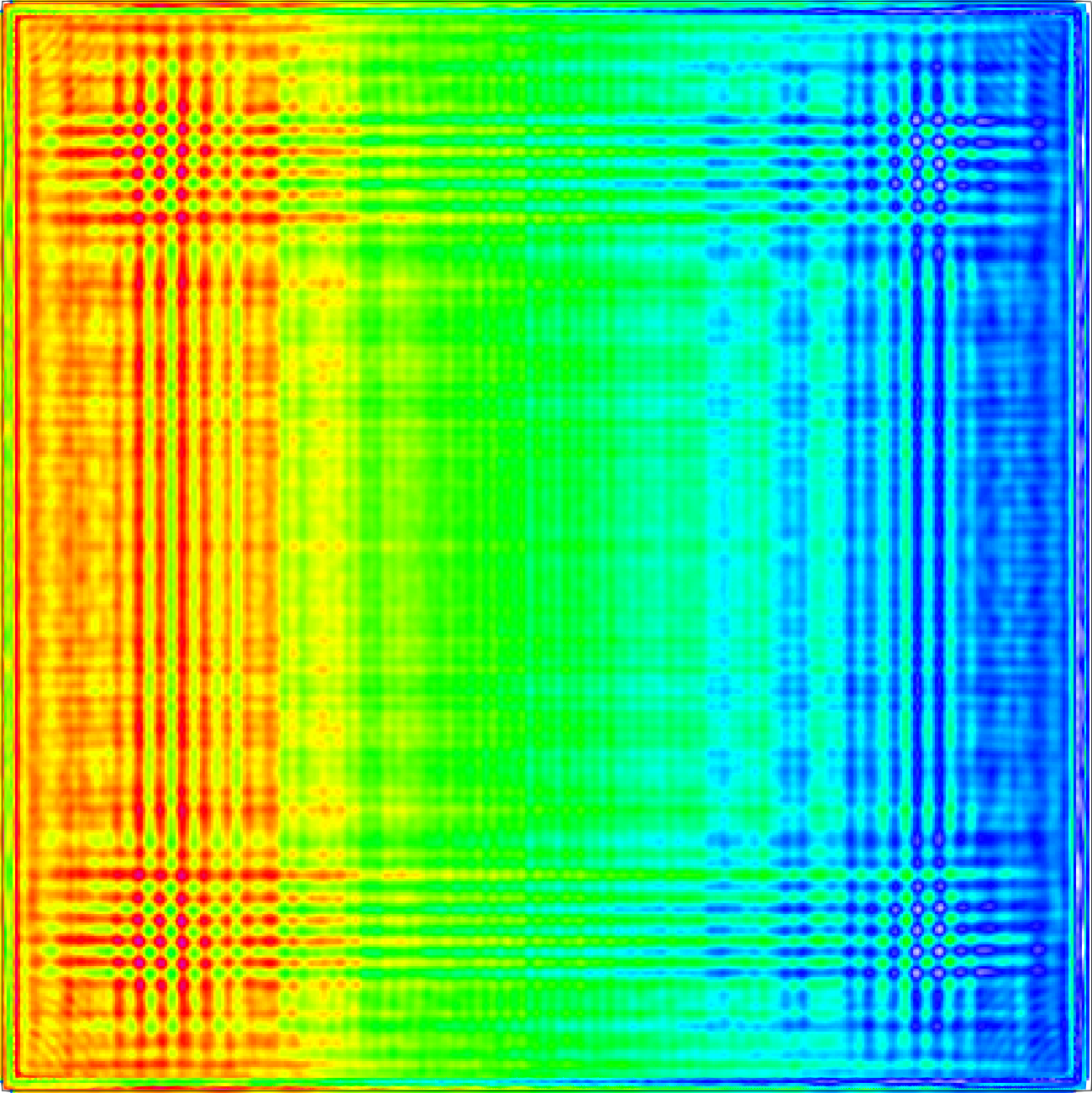}
    \input{adiabatic_expansion_snapshot_colorbar.pgf}
	\caption{Simulation snapshot at $t / T = 0.2$. The color bar represents pressure, in units of $\rho_0 c_0^2$Ma$^2$.}
	\label{fig:applications:adiabatic_expansion:snapshot}
\end{figure}

It should be incidentally remarked that this problem is exceptionally challenging due to the constant generation of pressure waves that are endlessly propagated within the fluid domain.
Those pressure waves are clearly seen in Fig. \ref{fig:applications:adiabatic_expansion:snapshot}, which shows the anisotropic pressure profile predicted above.

\begin{figure}[!ht]
	\centering
    \input{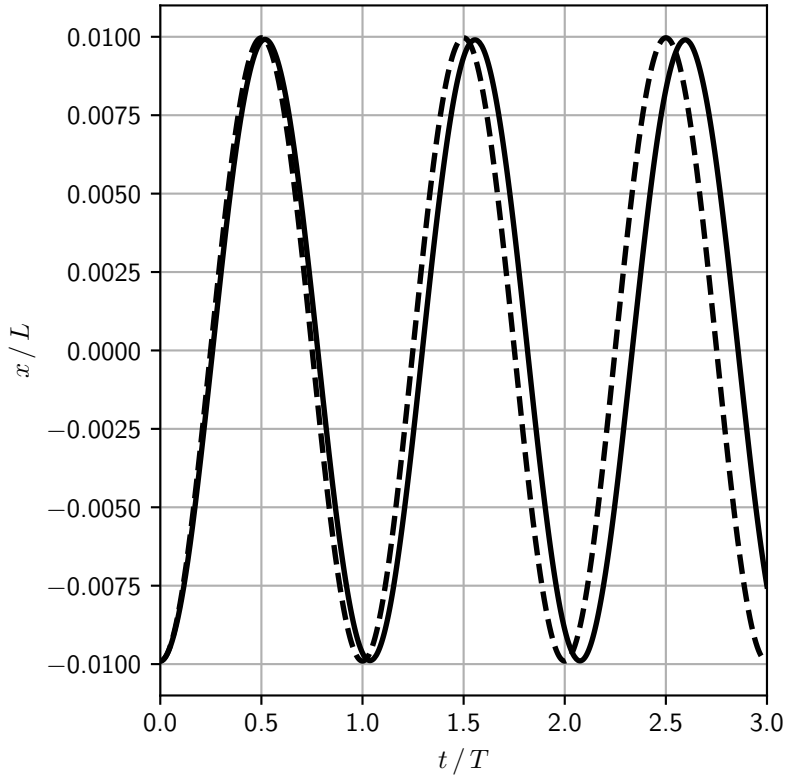}
	\caption{Position of the piston as a function of time in period units. Solid lines: simulation results; dashed line: harmonic approximation.
	}
	\label{fig:applications:adiabatic_expansion:x}
\end{figure}

\begin{figure}[!ht]
	\centering
    \input{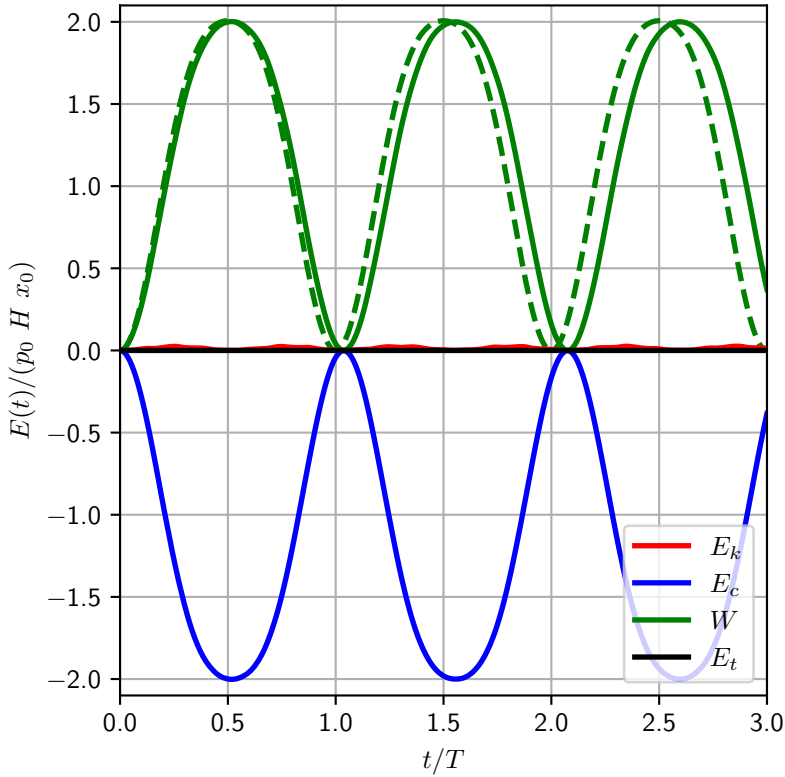}
    \caption{Energy of the fluid, in units of $p_0 H x_0$, as a function of time in period units. Green lines: energy transferred from the fluid to the piston; solid lines: simulation results; dashed line: harmonic approximation.
    Blue line: compression energy; red line: kinetic energy; black line: total energy.
    }
	\label{fig:applications:adiabatic_expansion:energy}
\end{figure}

In any case, the formulation described above is able to successfully perform the simulation, as can be appreciated in Figs. \ref{fig:applications:adiabatic_expansion:x} and \ref{fig:applications:adiabatic_expansion:energy}, where the position of the piston and the energy profile of the fluid are respectively depicted. These are compared with the results from the harmonic approximation (the results from solving the full equation of motion, Eq.~\eqref{eq:applications:adiabatic_expansion:EOM}).
%

\begin{figure}[!ht]
	\centering
    \input{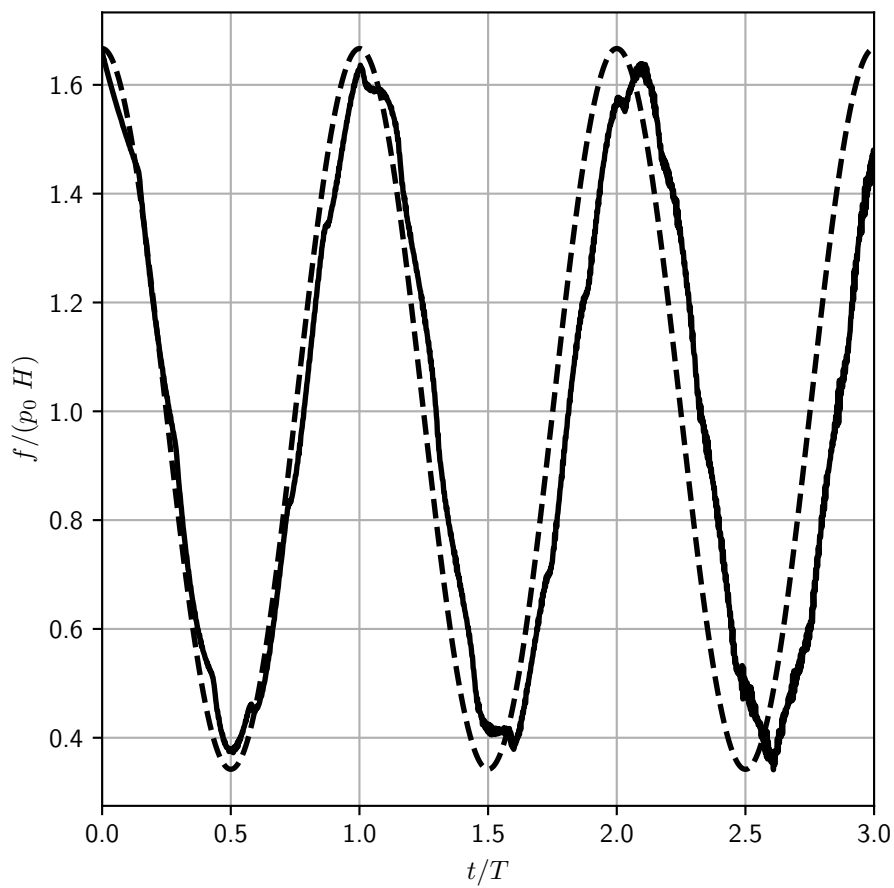}
	\caption{Forces on the piston, in units of $p_0 H $, as a function of time in period units. Solid black lines: simulation results, dashed black line: harmonic approximation.
	}
	\label{fig:applications:adiabatic_expansion:force}
\end{figure}

It is clear that the harmonic oscillator period is not met perfectly.
Along the same lines, it is also noticeable that the energy transferred to the wall during expansion is not perfectly returned to the fluid on contraction.
Anyway, an almost perfect total energy conservation is achieved, as well as a stable simulation.
The force is shown in Fig. \ref{fig:applications:adiabatic_expansion:force}: a reasonable approximation to the analytical force, with some higher frequency deviations due to acoustic waves.

\added{
As can be appreciated in the same Fig. \ref{fig:applications:adiabatic_expansion:force}, the force becomes noisier at the end of the simulation, reflecting the fact that the semi-implicit time scheme becomes increasingly challenging due to the unmitigated pressure waves mentioned above.
The position, energy profile, and the resulting forces are depicted in Figs.
\ref{fig:applications:adiabatic_expansion:x}, \ref{fig:applications:adiabatic_expansion:force}, and
\ref{fig:applications:adiabatic_expansion:energy}.
%
%
%
As can be seen, the force is again becoming noisier at the end of the simulation.
Effectively, the scheme is unconditionally stable, provided that the semi-implicit scheme is able to find a good solution at every time step.
}  
\subsection{Dam break}
\label{ss:applications:dam_break}
\begin{figure}[!ht]
	\centering
	\includegraphics[width=0.75\textwidth]{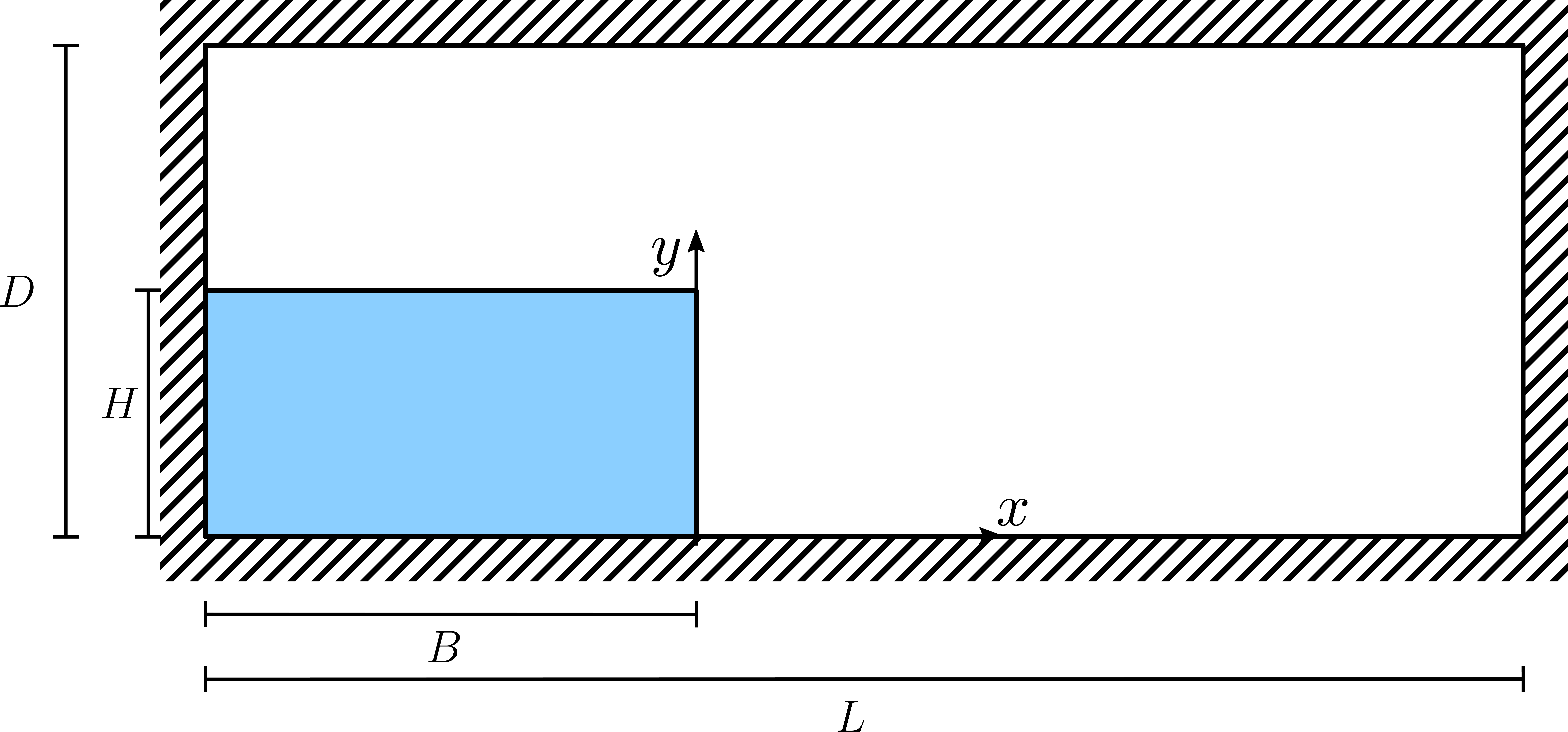}
	\caption{Schematic view of the dam break.}
	\label{fig:applications:dam_break:scheme}
\end{figure}

Another typical case that has been widely used in SPH to assess the quality of results in general, and energy conservation in particular, is the dam break (see for instance \cite{colagrossi_etal_pre2015_largestandingwave, zhang2021sphinxsys, antuono2021delta}).
A square reservoir of fluid of dimensions $B \times H$ is suddenly released to freely evolve in a tank of dimensions $L \times D$, as schematically depicted on Fig. \ref{fig:applications:dam_break:scheme}.

A detailed experimental investigation has been carried out on a tank of $L = 1.61\,  \mathrm{m}$ and $D = 0.6\,  \mathrm{m}$ \cite{lobovsky_etal_jfs2013_dambreak}.
In these experiments, a reservoir of water
($\rho_0 = 997\,  \mathrm{kg}\, \mathrm{m}^{-3}$,
$ \mu = 8.87 \times 10^{-4} \, \mathrm{Pa} \, \mathrm{s}$),
of dimensions $B = 0.6 \, \mathrm{m}$ and $H = 0.3 \, \mathrm{m}$ is considered.
The experiment was repeated 100 times, so good statistics on the impact pressure are obtained.

For the numerical investigation, the fluid reservoir is discretized with a particle inter-spacing of $\Delta x = H / 400$.
The same space between boundary elements is also considered.
A quintic Wendland kernel \cite{dehnen_aly_wendland_2012} is applied and its smoothing length $h$ is chosen so that $h / \Delta x = 2$.
The speed of sound of the fluid is calculated as
$c_0 = \sqrt{g H} / \text{Ma}$, with $g = 9.81 \mathrm{m s}^{-2}$ the acceleration of gravity and $\text{Ma} = 0.1$ the Mach number.
For the sake of a smoother simulation, the initial pressure is set to $p(\boldsymbol{r}, t=0) = \rho_0 g (H - \boldsymbol{r} \cdot \boldsymbol{j}) \cos(\pi \boldsymbol{r} \cdot \boldsymbol{i} / 2)$

The simulation is carried out with a semi-implicit midpoint time scheme, with $n_{\Delta t} = 10$ iterations per time step.
The midpoint semi-implicit iterator is stopped when the energy residue of Eq. (\ref{eq:applications:normal_impact:residue}) meets $R_{\Delta t}(t_{n+1/2}) < 10^{-6} \times E_p(t=0)$, with $E_p(t=0)$ the initial potential energy.
No relaxation is considered in the semi-implicit iterator at the beginning of each time step, although a progressively increasing relaxation factor is configured when the residue does not decrease at a geometric rate of $1/5$.
For time integration, a Courant number $\text{Co} = 0.1$ is imposed.

The energy profile depicted in
Fig. \ref{fig:applications:dam_break:energy} shows that the total
energy prior to the wall impact ($t \sqrt{g / H} \simeq 3$) is
almost perfectly conserved, with just $0.014\%$ dissipation, as
predicted by Ref. \cite{colagrossi_etal_pre2015_largestandingwave}.
\begin{figure}[!ht]
	\centering
    \input{dam_break_energy.pgf}
	\caption{Dam break: evolution of fluid energy.
    Red lines: kinetic energy,
    green: potential energy,
    blue: compression energy,
    black: total energy.}
	\label{fig:applications:dam_break:energy}
\end{figure}
After the impact, the dissipation increases faster, although the total energy dissipated at ($t \sqrt{g / H} \simeq 5$) is still rather low, just $0.35\%$.
At the end of the simulation, after the backsplash, the energy dissipation accelerates again, reaching $3.25\%$ at the end of the simulation.

\begin{figure}[!ht]
	\centering
	\includegraphics[width=0.95\textwidth]{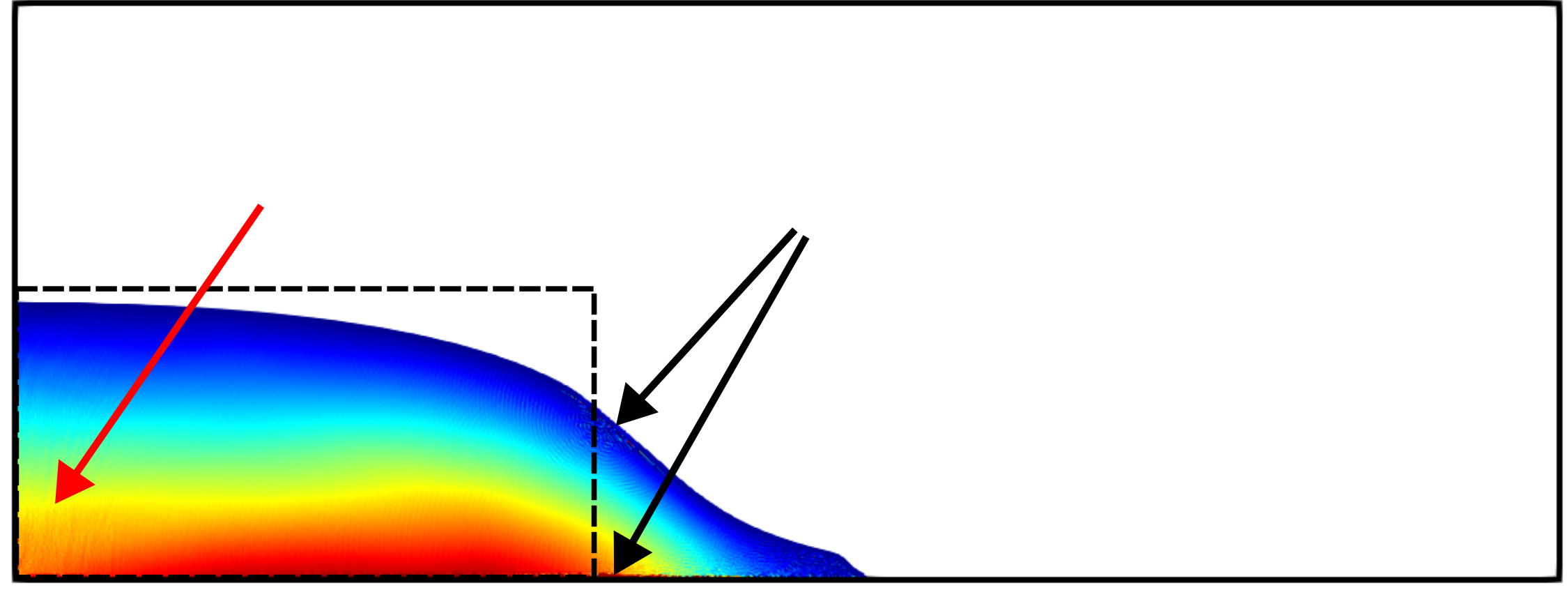} \\
	\includegraphics[width=0.95\textwidth]{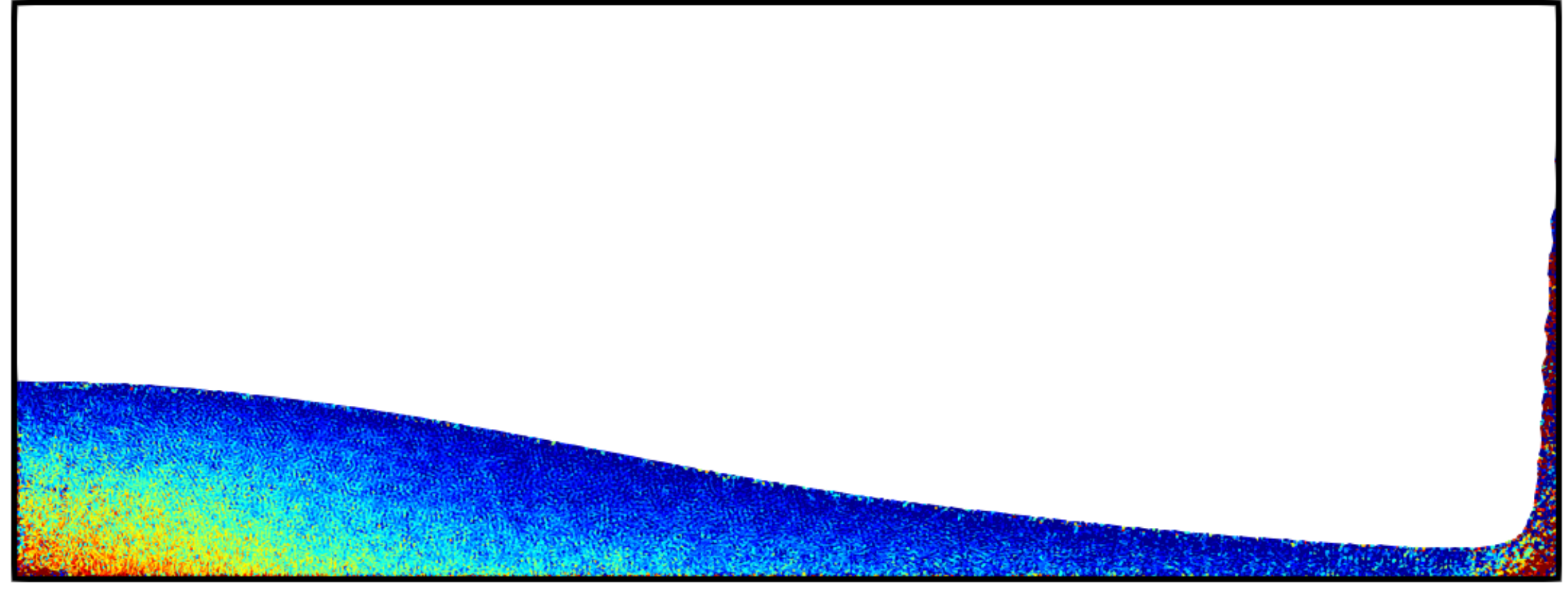} \\
    \input{dam_break_t3_colorbar.pgf}
	\caption{Snapshots of the simulation at times $t  \sqrt{g H}
          \simeq 1$ (top) and $ \simeq 3$ (bottom). The dashed black line denotes the initial shape of the fluid reservoir.
          Black arrows point to noise generated by the free surface
          inconsistencies, while red arrow points to the tensile
          instabilities.}
	\label{fig:applications:dam_break:snapshot}
\end{figure}

However, although the simulation becomes unconditionally stable, it is still afflicted with inconsistencies that generate some noise.
Indeed, as seen in the snapshots of Fig. \ref{fig:applications:dam_break:snapshot}, even if at the beginning of the simulation the pressure field is still quite smooth, inconsistencies arising from the free surface (see \cite{colagrossi_etal_jhr09}) and so-called tensile instabilities (see, for instance, \cite{monaghan2000sph}) can already be noticed.
Due to the lack of numerical dissipation, the noise generated is dissipated slower than it is generated, so at the time instant $t \sqrt{g / H} \simeq 3$ the pressure field looks noticeably noisy.
However, the simulation remains stable.

\begin{figure}[!ht]
	\centering
    \input{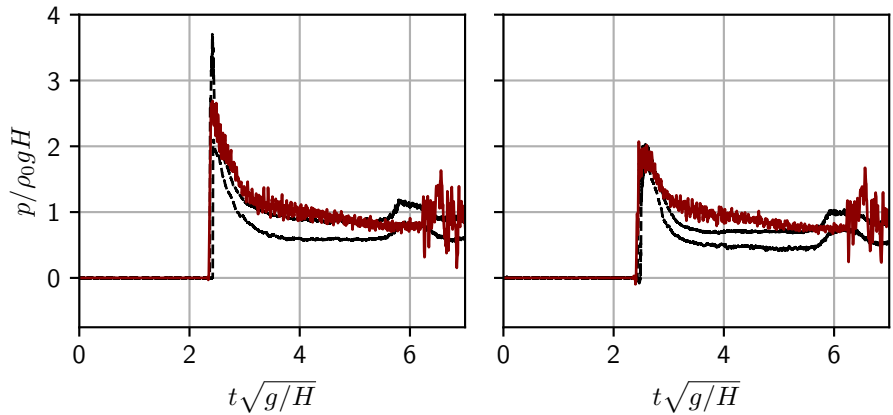}
	\caption{Pressure at sensor 1 (left) and sensor 2 (right). Black lines: experimental pressure median, 2.5\%, and 97.5\% percentiles. Red lines: SPH pressure. Deep red: smoothed SPH pressure.}
	\label{fig:applications:dam_break:sensors}
\end{figure}

Fig. \ref{fig:applications:dam_break:sensors} shows the pressure computed at sensors 1 and 2.
The pressure peaks are nicely captured, although the tail of the pressure curve is overestimated by SPH.
This is in contrast with Ref. \cite{antuono2021delta}, where the pressure peak is underestimated, while the tail is captured quite well.
However, the differences are consistent with the fact that the $\delta$-ALE-SPH model considered in Ref. \cite{antuono2021delta} implies numerical dissipation.

\begin{figure}[!ht]
	\centering
	\includegraphics[width=0.9\textwidth]{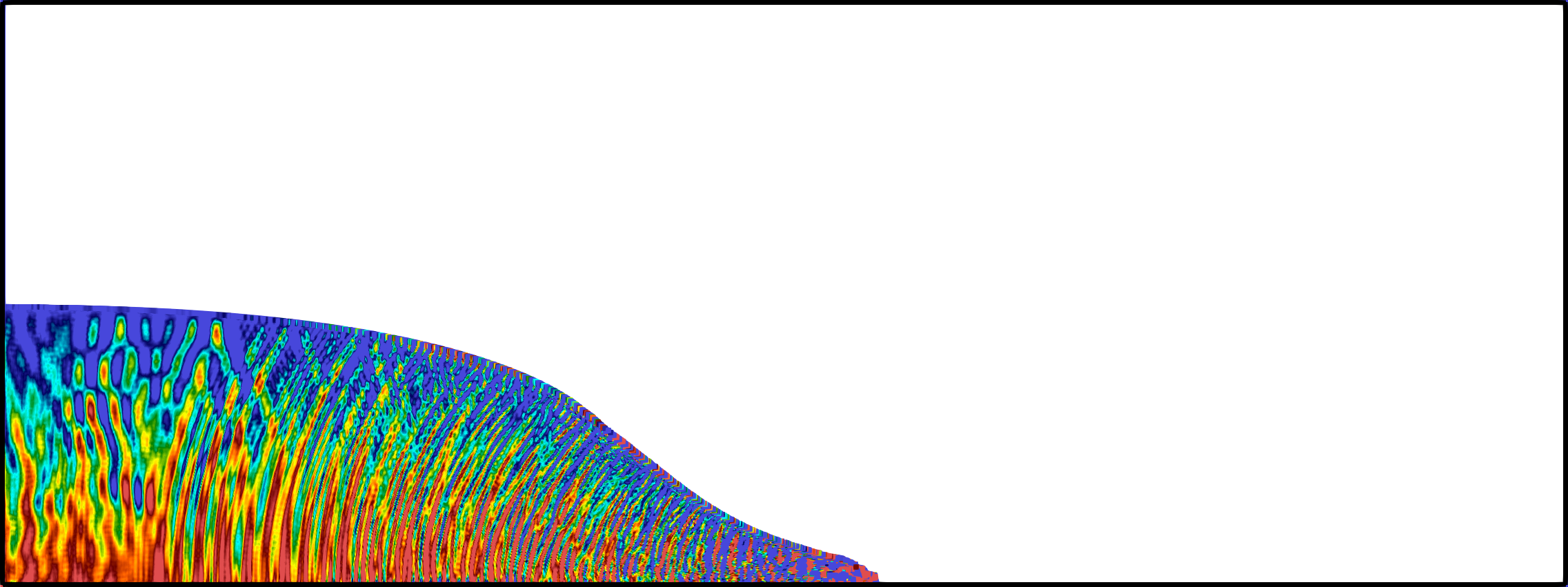} \\
    \input{dam_break_t1_bi_colorbar.pgf}
	\caption{Snapshot of the simulation considering plain BI. $t  \sqrt{g / H} \simeq 1$}
	\label{fig:applications:dam_break:snapshot_bi}
\end{figure}

For completeness, the same simulation is carried out considering BI, as introduced in Eqs.~(\ref{eq:sph:boundaries:bi:gradp}, \ref{eq:sph:boundaries:bi:divu}).
As can be seen in the snapshot in Fig. \ref{fig:applications:dam_break:snapshot_bi}, at an early stage of the simulation, $t  \sqrt{g / H} = 1$, the pressure profile is heavily affected by noise.
The pressure waves are very clearly seen in this snapshot.
The energy profile for this BC alternative is depicted in Fig. \ref{fig:applications:dam_break:energy_bi}, where
the amount of extra energy pumped onto the system is shown to be relatively small with this method, about $0.25\%$ of the initial potential energy.
\begin{figure}[!ht]
	\centering
    \input{dam_break_energy_bi.pgf}
	\caption{
    Dam break with BI, evolution of fluid energy.
    Red lines: kinetic energy,
    green: potential energy,
    blue: compression energy,
    black: total energy.
 }
	\label{fig:applications:dam_break:energy_bi}
\end{figure}
\added{%
It is nevertheless large enough to make the simulation unstable.
}

\begin{figure}[!ht]
	\centering
    \input{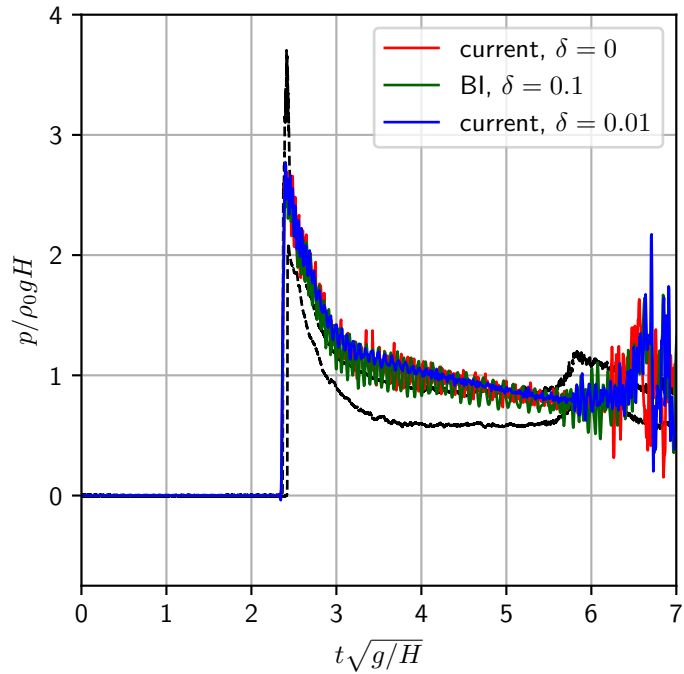}
	\caption{\added{Pressure at sensor 1 for different formulations. Black lines: experimental pressure median, 2.5\%, and 97.5\% percentiles.}}
	\label{fig:applications:dam_break:p_comp}
\end{figure}

\added{%
Although the scheme proposed in this manuscript is unconditionally stable, incorporating dissipative methodologies \cite{vila1999, antuono2015energy, CercosPita2016} can further enhance the results without significantly increasing computational costs.
Fig. \ref{fig:applications:dam_break:p_comp} compares the pressure signals at the same sensor for two boundary conditions, those described in this manuscript, and plain BI \cite{ferrand_etal_2012} when $\delta$-SPH is applied.
}
\added{%
Notably, a small amount of $\delta$ dissipation significantly reduces the signal noise with the new formulation.
However, with plain BI this $\delta$ dissipation value must be increased tenfold to achieve a comparable reduction.
Moreover, the new formulation does not require any parameter tuning, given its unconditional stability.
In contrast, other boundary conditions require the balance of the artificial energy pumped into the system with the extra dissipation provided by $\delta$-SPH.
}

\begin{figure}[!ht]
	\centering
    \input{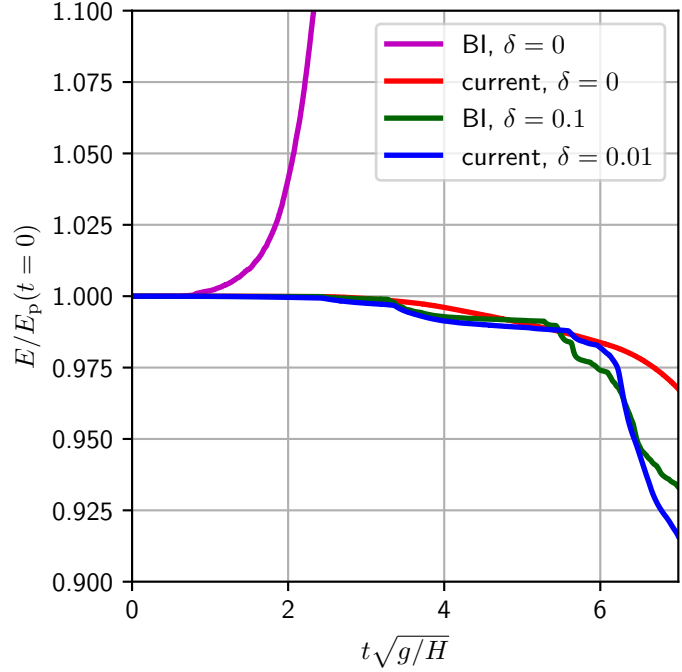}
	\caption{\added{Total energy along the simulation, considering different formulations.}}
	\label{fig:applications:dam_break:e_comp}
\end{figure}

\added{%
Fig. \ref{fig:applications:dam_break:e_comp} illustrates the total energy over the course of the simulation for all formulations considered.
}
\added{%
While the new formulation does not perfectly conserve total energy (resulting in a 3.25\% energy loss by the end of the simulation, after the backsplash) it still conserves energy far better than the other formulations.
In fact, plain BI leads to a divergent result without additional dissipation, while applying $\delta$-SPH stabilizes it.
However, this stabilization comes at the cost of increased energy dissipation: 8.39\% for the new formulation and 6.72\% for the BI.
Incidentally, it should be noted that some researchers have reported situations where the application of $\delta$-SPH on its own was not enough to stabilize the simulations (see, e.g., \cite{martinez_carrascal_2023_sloshing}.)
}

\subsection{Spacecraft water landing}
\label{ss:applications:apollo_capsule}
\added{
Undoubtedly, one of the key features that makes SPH particularly attractive to CFD practitioners is its ability to handle complex geometries, as demonstrated in various practical applications (see, among others, \cite{Marrone2011, Adami20127057, lyu_2024_wading_car}).
}

\added{%
This application is a simulation of the experiments conducted at the Langley Research Center on an Apollo capsule landing on water \cite{stubbs_1967_apollo}.
While these experiments have previously been addressed using SPH (see, for instance, \cite{lu_2017_apollo}), our target is less ambitious:
rather than covering the full set of experiments, we focus on assessing the capabilities of the new formulation in handling highly curved complex geometries on a violent impact case.
Therefore, only one pitch angle is considered here.
}

\begin{figure}[!ht]
	\centering
	\includegraphics[width=0.6\textwidth]{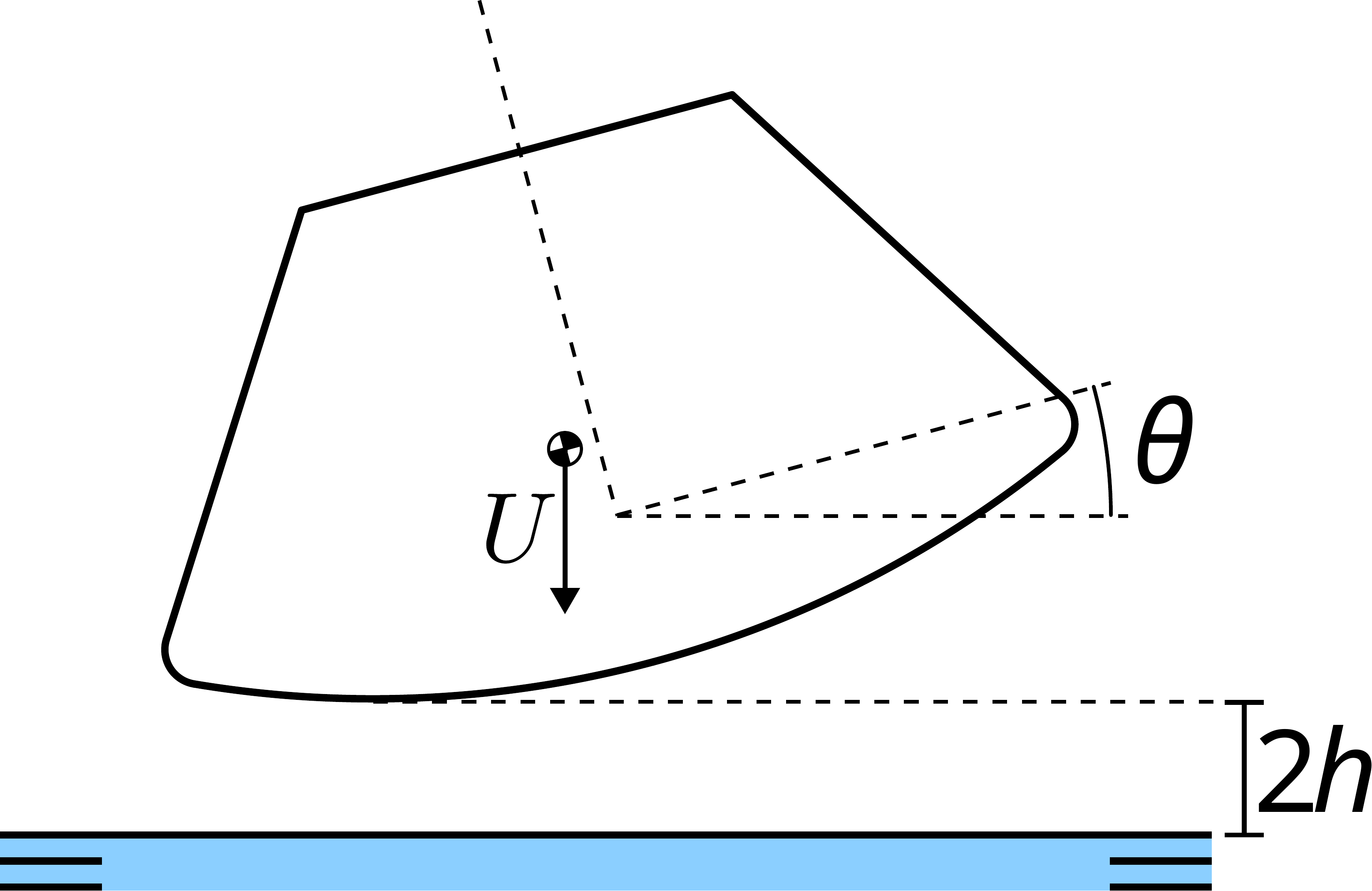}
	\caption{\added{%
			Schematic view of the initial condition for the spacecraft water landing.
		}
        }
	\label{fig:applications:apollo_capsule:scheme}
\end{figure}

\added{%
The simulation consists of an axially symmetric spacecraft landing on water with a prescribed initial pitch angle and velocity.
A schematic view of the initial condition is shown in Fig. \ref{fig:applications:apollo_capsule:scheme}.
For further details on dimensions, mechanical properties, and initial configurations, we refer the reader to the original publication \cite{stubbs_1967_apollo}.
}

\added{%
In the initial condition, the spacecraft is positioned at a distance $2h$ from the water free surface, with an initial pitch angle $\theta = 12 \, \mathrm{deg}$ and downward vertical velocity of $U = 9.08 \, \mathrm{m/s}$.
The water tank is modeled as a box with dimensions $L = B = 16 \, \mathrm{m}$ and $H = 8 \, \mathrm{m}$, filled with water that has a reference density of $\rho_0 = 998 \, \mathrm{kg/m}^3$.
The fluid is discretized using a regular lattice of particles with three different inter-spacing values: $\Delta x = 0.32 \, \mathrm{m}$, $\Delta x = 0.16 \, \mathrm{m}$, and $\Delta x = 0.08 \, \mathrm{m}$.
A quintic Wendland kernel \cite{dehnen_aly_wendland_2012} is applied with a smoothing length $h  = 2  \Delta x$.
The speed of sound in the fluid is set to values $c_0 = 5 \times 10^2 \, \mathrm{m/s}$, $c_0 = 10^3 \, \mathrm{m/s}$ and $c_0 = 1.5 \times 10^3 \, \mathrm{m/s}$.
The pressure field is initialized as the hydrostatic one, $p(\boldsymbol{r}, t=0) = \rho_0 g (H - \boldsymbol{r} \cdot \boldsymbol{k})$.
}

\added{%
The simulation is carried out with a semi-implicit midpoint time scheme, with $n_{\Delta t} = 10$ iterations per time step.
The midpoint semi-implicit iterator is stopped when the energy residue of Eq. (\ref{eq:applications:normal_impact:residue}) meets $R_{\Delta t}(t_{n+1/2}) < 5 \times 10^{-8} \times E_k(t=0)$, with $E_k(t=0)$ the initial kinetic energy of the spacecraft.
No relaxation is used in the semi-implicit iterator.
For time integration, several Courant numbers are tested: $\text{Co} = 0.1$, $\text{Co} = 0.25$ and $\text{Co} = 0.5$.
}

\begin{figure}[!ht]
	\centering
	\includegraphics[width=0.7\textwidth]{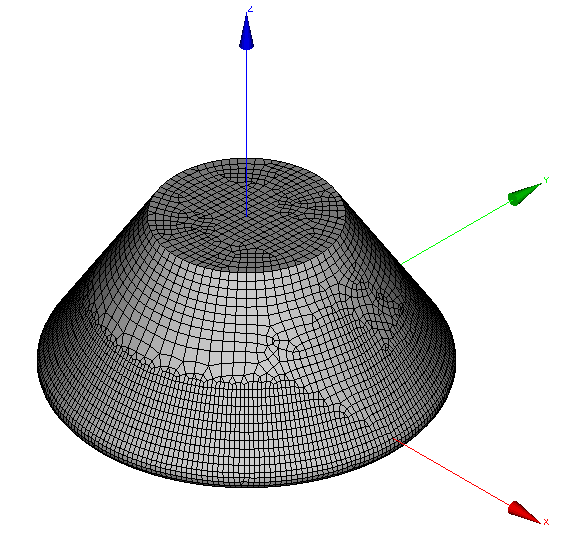}
	\caption{\added{
        Snapshot of the mesh generated for the Apollo spacecraft model.
    }
    }
	\label{fig:applications:apollo_capsule:mesh}
\end{figure}

\added{%
Regarding the boundary elements, the walls of the water tank are discretized using 2D regular lattices with the same inter-spacing as the fluid.
In contrast, the spacecraft is modeled using the free and open-source FreeCAD package \cite{riegel_2016_freecad}, and meshed with the Salome open-source package 
\cite{salomeWebSite}%
, employing a quadrangular-dominated mesh generated by Netgen \cite{schoberl_2009_netgen}.
The edge length of the mesh is limited to be between $\Delta x = 0.08 \, \mathrm{m}$ and $\Delta x = 0.04 \, \mathrm{m}$.
A snapshot of the generated mesh is shown in Fig. \ref{fig:applications:apollo_capsule:mesh}.
It should be noted that the generated mesh does not exhibit any symmetry.
Boundary elements are placed at the centroid of each face of the mesh with the appropriate face area.
}

\added{%
After the initial condition, the spacecraft is left to evolve freely.
To model the rigid body dynamics the Free-Software AQUAgpusph \cite{CercosPita2015} used in this work is loosely coupled with the open-source library Project Chrono 
\cite{projectChronoWebSite}%
.
Specifically, a staggered simulation approach is adopted: after executing a time step in AQUAgpusph, the forces on the rigid body are collected. These forces are then applied in a Project Chrono time step, which updates the position and velocity of the rigid body. The updated state is used subsequently for the next AQUAgpusph time step.
It should be noticed that the rigid body dynamics are thus not included in the semi-implicit iterative scheme, hence a mismatch in energy transfer between the fluid and the rigid body can be expected.
}

\begin{figure}[!ht]
	\centering
    \input{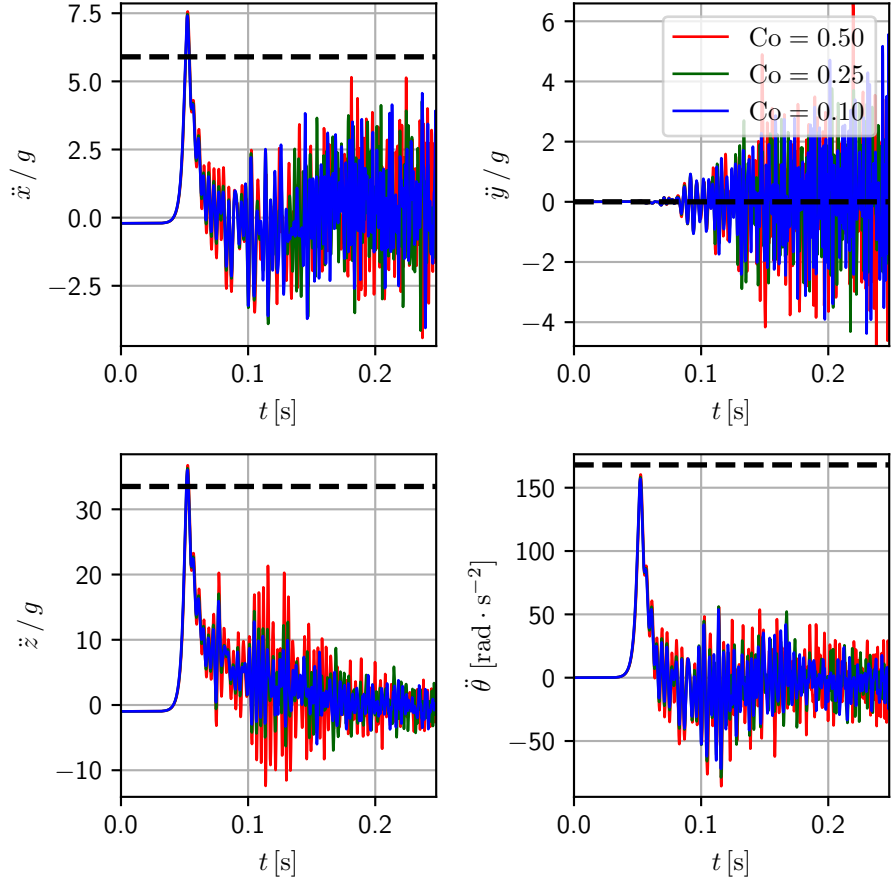}
	\caption{\added{Accelerations on the spacecraft for
			varying Courant numbers and
			 fixed resolution $\Delta x = 0.16 \, \mathrm{m}$ and speed of sound $c_0 = 1000 \, \mathrm{m / s}$.
		Red lines: Co$=0.1$,
		green lines: Co$=0.25$,
		blue lines: Co$=0.5$.
	    Dashed line: Acceleration measured during the experiments.
	}
    }
	\label{fig:applications:apollo_capsule:acc_co}
\end{figure}

\added{%
In contrast with some previous results\cite{lu_2017_apollo}, the time step, determined by the Courant number, does not appear to affect significantly the outcome.
In fact, the accelerations shown in Fig. \ref{fig:applications:apollo_capsule:acc_co}
(with fixed spatial resolution $\Delta x = 0.16 , \mathrm{m}$ and speed of sound $\Delta c_0 = 1000 , \mathrm{m/s}$) show a minimal impact of the Courant number.
}

\added{%
Overall, the simulation behaves as expected.
After a short flying time, the rigid body begins to feel the effect of the water surface, quickly reaching the maximum acceleration, due to the spacecraft slamming into the water.
Following this, the acceleration quickly decreases as the capsule slowly sinks and rotates.
}

\added{%
Noticeable acceleration values are registered along the $y$ axis, caused by the asymmetric disposition of the boundary elements. However, this signal is basically white noise, resulting in negligible velocities for the capsule along this direction.
}

\begin{figure}[!ht]
	\centering
    \input{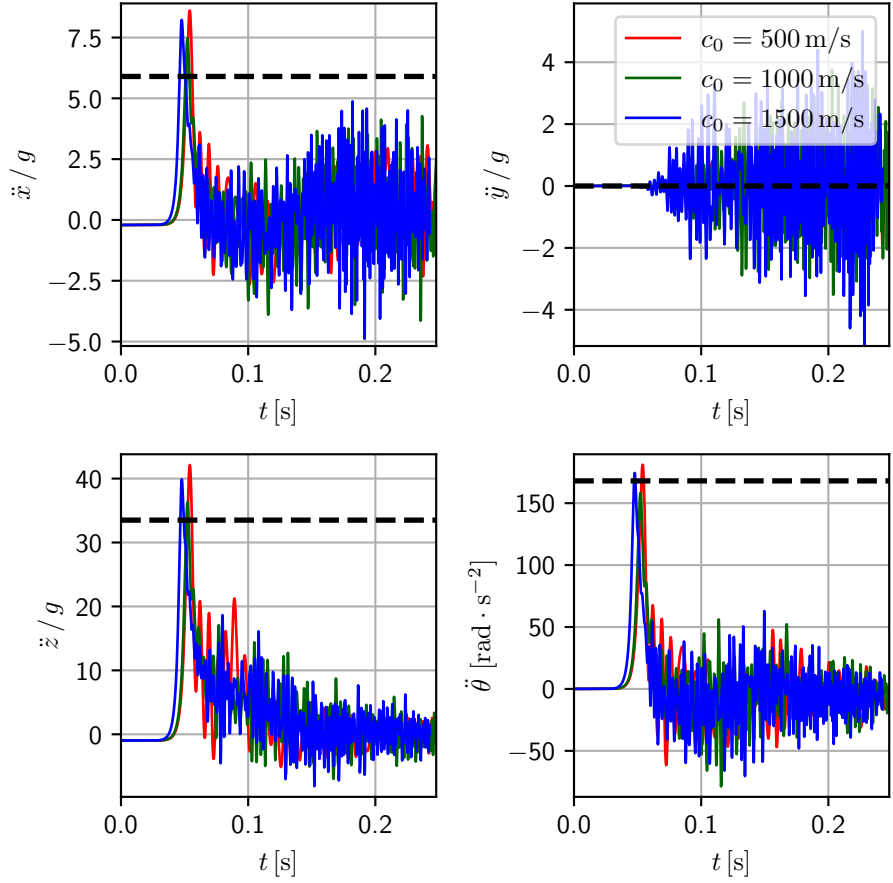}
	\caption{\added{Computed accelerations on the spacecraft for
			varying speeds of sound and fixed resolution $\Delta x = 0.16 \, \mathrm{m}$  and Courant number $\text{Co} = 0.25$.
    	Red lines: $c_0=500$~m/s,
    	green lines: $c_0=1000$~m/s,
    	blue lines: $c_0=1500$~m/s.
    	Dashed line: Acceleration measured during the experiments.}
    }
	\label{fig:applications:apollo_capsule:acc_cs}
\end{figure}

\begin{figure}[!ht]
	\centering
    \input{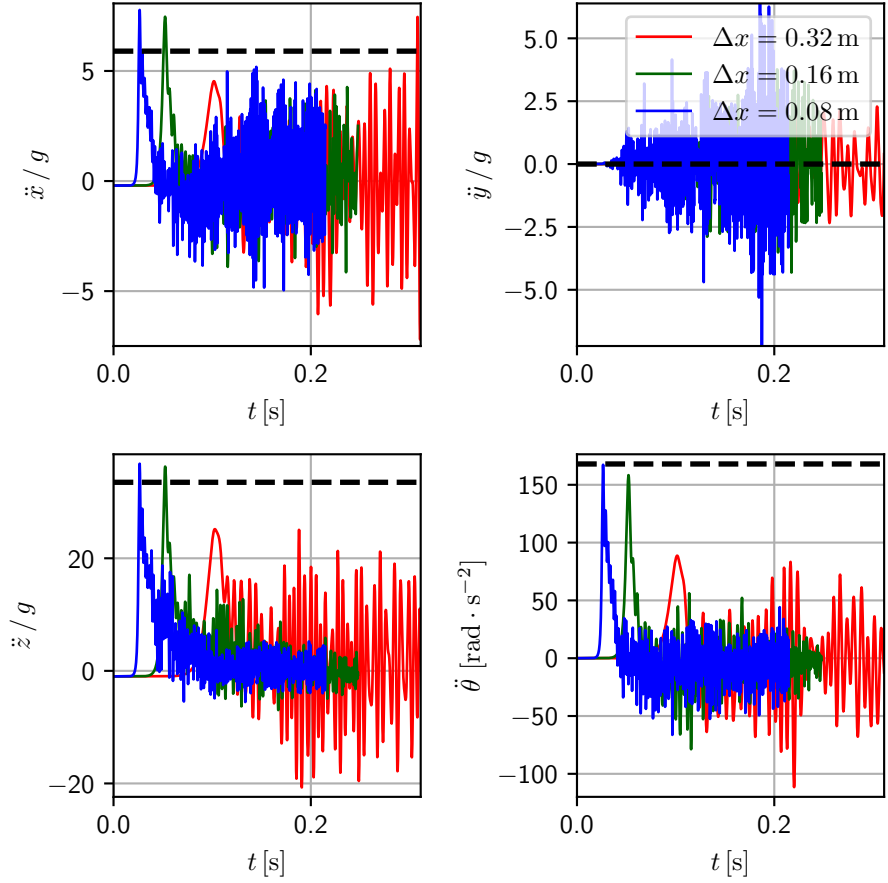}
	\caption{\added{Computed accelerations on the spacecraft for
			varying resolutions and a fixed speed of sound
			$c_0 = 1000 \, \mathrm{m/s}$  and Courant number $\text{Co} = 0.25$.
		Red lines: $\Delta x=0.08$~m,
		green lines: $\Delta x=0.16$~m ,
		blue lines: $\Delta x=0.32$~m.
		Dashed line: Acceleration measured during the experiments.}
    }
	\label{fig:applications:apollo_capsule:acc_dr}
\end{figure}

\added{%
Although the time step does not play a critical role in the accelerations, the same cannot be said about the speed of sound and the spatial resolution, as shown in Figs. \ref{fig:applications:apollo_capsule:acc_cs} and \ref{fig:applications:apollo_capsule:acc_dr}.
}


\added{%
The speed of sound has a noticeable impact on the results.
However, for the speeds of sound considered, the results seem to have already converged.
The computed accelerations tend to over-predict the linear accelerations measured during the experiments.
However, it should be noted that the maximum acceleration is over-predicted for a time lapse on the order of 1 ms, which is 66\% shorter than the accelerometer sampling frequency.
}

\added{%
Regarding the resolution, it exhibits a different trend.
Using a coarse resolution, $\Delta x = 0.32 \, \mathrm{m}$, the acceleration results clearly mismatch those measured during the experiments.
However, the results converge rather quickly.
Indeed, the differences between the two finest resolutions are barely noticeable, except for the length of the flying time.
}

%
%
%

\begin{figure}[!ht]
	\centering
	\input{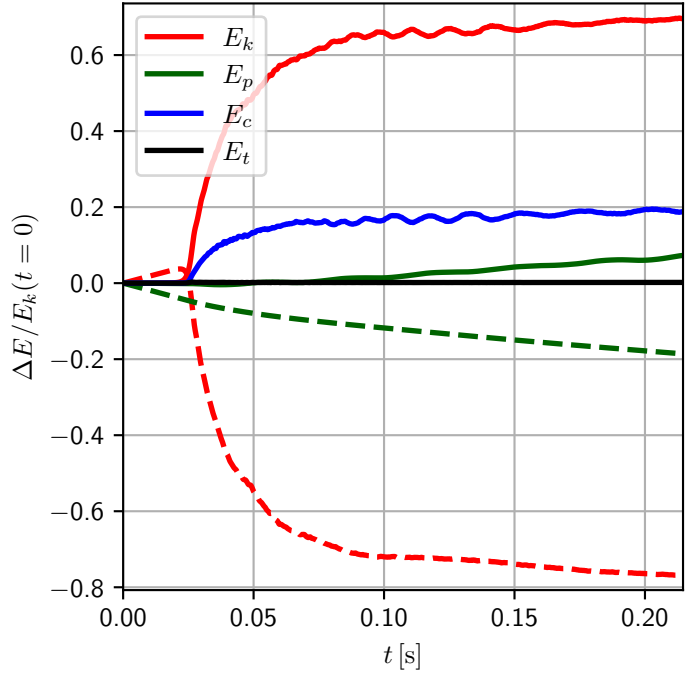}
	\caption{\added{Energy profile with
			 $\Delta x = 0.16 \, \mathrm{m}$,
			 $\Delta c_0 = 1000 \, \mathrm{m / s}$,
			 and $\text{Co} = 0.25$.
    	    Solid colored lines: Fluid energy components.
        	Dashed colored lines: spacecraft energy components.
        	Solid black line: Sum of all the energy components
    }
    }
	\label{fig:applications:apollo_capsule:e}
\end{figure}

\added{%
Regarding the energy evolution, all the components of the energy are plotted in Fig. \ref{fig:applications:apollo_capsule:e}, for the finest resolution case: $\Delta x = 0.16 \, \mathrm{m}$, $\Delta c_0 = 1000 \, \mathrm{m / s}$ and $\text{Co} = 0.25$.
}

\added{%
As can be seen, during the flight time the capsule gains a small amount of kinetic energy, which is largely transferred to the fluid upon the violent impact.
Throughout the entire simulation, net energy is transferred from the spacecraft to the fluid.
However, in a longer simulation, part of this energy would be transferred back to the solid.
}

\added{%
It should be highlighted that, despite the large considered speed of sound, a significant amount of the fluid energy is stored in the compressible component.
}

\begin{figure}[!ht]
	\centering
    \input{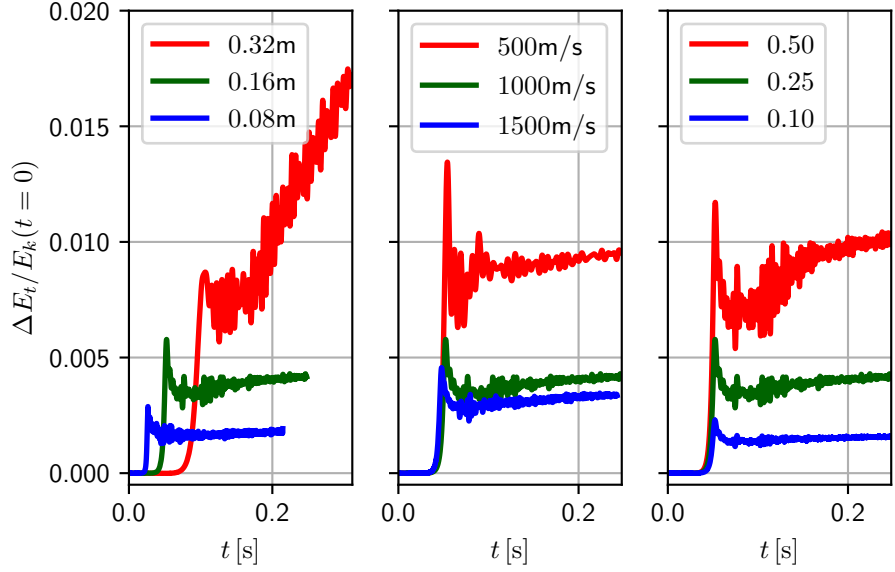}
	\caption{\added{Total energy profile for varying
			Courant numbers (left), resolution (center),
			and speeds of sound (right).
			Left: fixed resolution $\Delta x = 0.16 \, \mathrm{m}$
			and speed of sound $c_0 = 1000 \, \mathrm{m / s}$.
            Center: fixed speed of sound $c_0 = 1000 \, \mathrm{m / s}$
			and Courant number $\text{Co} = 0.25$.
			Right: fixed resolution $\Delta x = 0.16 \, \mathrm{m}$
			and Courant number $\text{Co} = 0.25$.
    }
    }
	\label{fig:applications:apollo_capsule:e_co_dr_cs}
\end{figure}

\added{%
As mentioned earlier, the net energy transferred from the solid to the fluid is not perfectly conserved due to the loose coupling considered. Fig.
\ref{fig:applications:apollo_capsule:e_co_dr_cs} shows the total energy profiles obtained by varying one parameter while keeping the other two fixed.
As shown, the total energy is never perfectly conserved.
Moreover, across all simulations, there is a net generation of extra energy,
which decreases as the simulation parameters are refined, that is, with a smaller time step, a larger speed of sound, or a finer resolution.
Quantitatively, for the coarsest resolution, the amount of spuriously generated energy is less than 1.75\% of the initial kinetic energy, while for the finest resolution it decreases to 0.25\%.
}

%
%

\begin{figure}[!ht]
	\centering
	\includegraphics[width=0.98\textwidth]{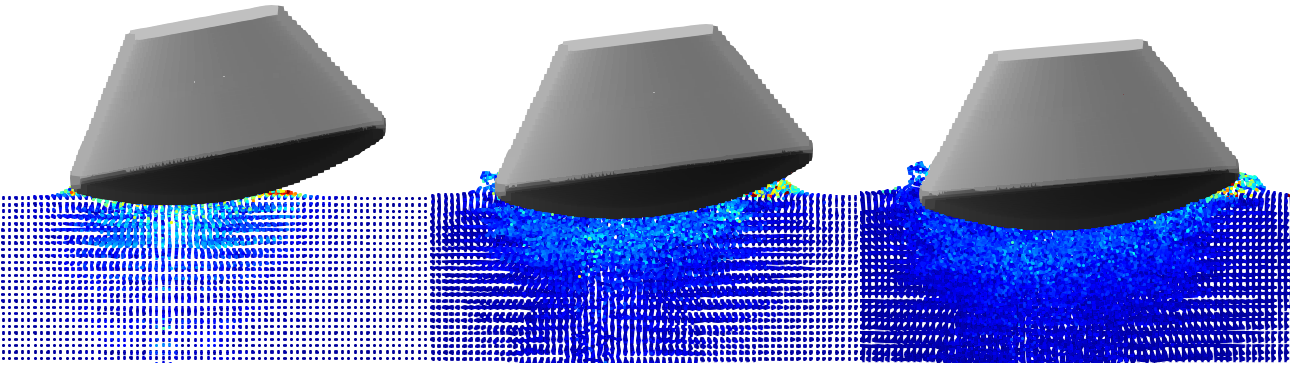} \\
	\input{apollo_capsule_snapshot_Ma_colorbar.pgf}
	\caption{\added{Detail on the particle velocities with parameters as in
			Fig.~\ref{fig:applications:apollo_capsule:e}.
			Left: $t = 0.05 \, \mathrm{s}$. Center: $t = 0.1 \, \mathrm{s}$. Right: $t = 0.15 \, \mathrm{s}$. The capsule is represented by the shaded boundary elements. The represented scenes are clipped by the $y=0$ plane
  	  }
    }
	\label{fig:applications:apollo_capsule:snapshots_Ma}
\end{figure}

\begin{figure}[!ht]
	\centering
	\includegraphics[width=0.98\textwidth]{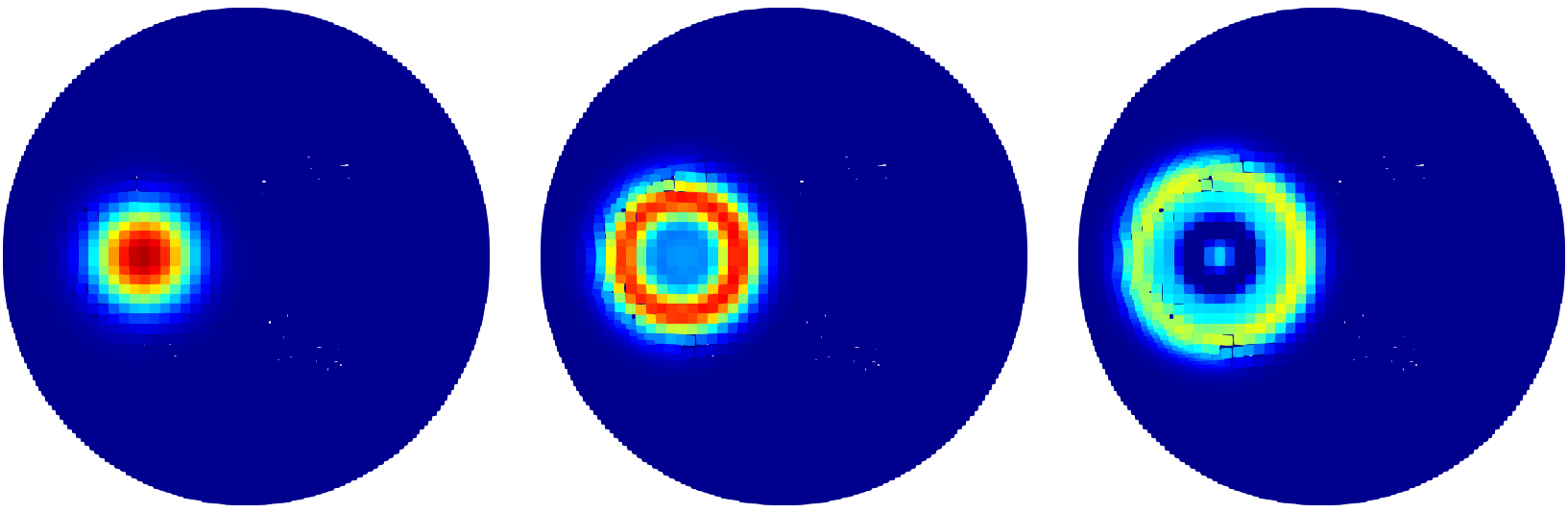} \\
	\input{apollo_capsule_snapshot_p_colorbar.pgf}
	\caption{\added{Detail on the pressure over the capsule bottom surface,
			with parameters as in Fig.~\ref{fig:applications:apollo_capsule:e}.
			Left: $t = 0.025 \, \mathrm{s}$. Center: $t = 0.027 \, \mathrm{s}$. Right: $t = 0.029 \, \mathrm{s}$
    }
    }
	\label{fig:applications:apollo_capsule:snapshots_p}
\end{figure}

\added{%
For completeness, in Fig. \ref{fig:applications:apollo_capsule:snapshots_Ma} several snapshots of the simulation are shown.
As can be seen, no particles are trespassing the boundary.
The pressure on the spacecraft surface during the impact is detailed in Fig.\ref{fig:applications:apollo_capsule:snapshots_p} showing
large pressures being created at the contact point and then moving out at the speed of sound.
}

\section{Conclusions and future work}
\label{s:conclusions}
A novel wall boundary condition that ensures exact energy conservation is here introduced. \added{This proposal is formulated within the framework of boundary integrals, but it is different from previous BI formulations such as our previous
article, Ref. \cite{cercospita_2023_energy}.}
Its main advantage is the unconditional stability of the resulting model, without the need for any form of dissipation, either numerical or physical.
However, a notable limitation is the relatively low order of consistency, particularly when compared to previous formulations.
Nonetheless, the new formulation demonstrates excellent performance in the numerical tests, reducing wall penetrability and providing accurate results.

Regrettably, certain tasks remain unaddressed in this publication.
One of the future targets will be to explore higher-order consistency operators to enhance the model.
Additionally, consistency issues arising from the so-called tensile instability will be tackled by a particle-shifting formulation that upholds energy conservation.
\section*{CRediT authorship contribution statement}
Jose Luis Cercos-Pita: Methodology, Formal analysis, Software, Writing

Daniel Duque: Methodology, Formal analysis, Writing

Pablo Eleazar Merino-Alonso: Methodology, Software

Javier Calderon-Sanchez: Methodology, Software
\section*{Declaration of competing interest}
The authors declare that they have no known competing financial interests or personal relationships that could have influenced the work herein reported.
\section*{Data availability}
Data will be made available upon reasonable request. The code
can be found in the \textit{AQUAgpusph} gitlab repository.

\section*{Acknowledgments}
The authors acknowledge support from
projects
\textit{TED2021-130951B-I00, FOWT-PLATE-MOOR: Hidrodinámica de heave-plates y líneas de fondeo para turbinas eólicas flotantes}, and
\textit{PID2021-123437OB-C21,
UPM-FOWT-DAMP2: Hidrodinámica de Elementos de Amortiguamiento del Movimiento de Aerogeneradores Flotantes: Modelo numérico y experimentos a pequeña escala},
both from Spain's Ministerio de Ciencia e Innovación.

\appendix
\section{Consistency notes on Boundary Integrals}
\label{s:bi_approx}

The approximation that leads to BI can be written on a more formal way as
\begin{equation}
\Grad{f}(\bs{y}) = \SPHint{\Grad{f}}(\bs{x}) + O(h) O(\Lap{f}(\bs{x})) + O(h^2) \quad \forall \bs{y} \in \bar{\Omega},
\end{equation}
which leads to the following expression:
\begin{equation} \label{eq:bi_approx:general}
\begin{split}
- \frac{1}{\gamma(\bs{x})} \int_{\Omega} f(\bs{y}) \Grad{W}(\bs{y} - \bs{x}) \D \bs{y} + \frac{1}{\gamma(\bs{x})} \int_{\Omega} f(\bs{y}) \bs{n}(\bs{y}) W(\bs{y} - \bs{x}) \D \bs{y} = \\
\SPHint{\Grad{f}}(\bs{x}) + O(h) O(\Lap{f}(\bs{x})) + O(h^2).
\end{split}
\end{equation}

Thus, within the BI formulation, as described in Refs. \cite{ferrand_etal_2012, cercospita_thesis_2016}, the resulting gradient is just an approximation of the expected weak value, $\SPH{\Grad{f}}(\bs{x})$.
Nevertheless, a BI formulation may be consistent for the computation of first order differential operators --- 
indeed, formulations that fulfill the conditions of Eqs.~\eqref{eq:sph:boundaries:gp:gradp_consistency} have been proposed  \cite{Maciaetal_PTP_2012}.
\section{Energy conservation for infinite symmetry planes}
\label{s:sym_plane}

Let us consider an infinite symmetry plane featured by its outward pointing normal, tangent and binormal vectors: $\bs{n}$, $\bs{t}$, and $\bs{b} = \bs{n} \times \bs{t}$.
Also, for the sake of simpler notation, the reference is set at some point of the plane, that is, $\bs{x} \cdot \bs{n} = 0 \quad \forall \bs{x} \in \partial \Omega$.

Now assume, as is common practice on symmetry planes, that the extended fluid is modeled by mirroring the fluid particles.
Thus, each arbitrary $i$-th fluid particle has a mirrored extended $i'$-th particle, with the following properties:
\begin{align}
\bs{r}_{i'} = \bs{r}_{i} + 2 (\bs{r}_{i} \cdot \bs{n}) \bs{n},
\\
\bs{u}_{i'} = \bs{u}_{i} + 2 (\bs{u}_{i} \cdot \bs{n}) \bs{n},
\\
\rho_{i'} = \rho_i.
\\
p_{i'} = p_i.
\end{align}

Hence, Eq. \eqref{eq:energy:homogeneous:gp} can be rewritten as
\begin{equation}
\SPH{P}^{\partial \Omega} = \sum_{i \in \Omega} \sum_{j \in \Omega} \frac{m_i}{\rho_i} \frac{m_j}{\rho_j} \left(p_j \bs{u}_i + p_i \bs{u}_{j'} \right) \cdot \Grad{W}_{ij'},
\end{equation}
with $\Grad{W}_{ij'} = \Grad{W}\left(\bs{r}_{j'} - \bs{r}_{i}\right)$.
All terms may be projected into normal and tangent directions,
\begin{align}
\SPH{P}^{\partial \Omega}_n = \sum_{i \in \Omega} \sum_{j \in \Omega} \frac{m_i}{\rho_i} \frac{m_j}{\rho_j} \left(p_j \bs{u}_i \cdot \bs{n} - p_i \bs{u}_{j} \cdot \bs{n} \right) \left( \Grad{W}_{ij'} \cdot \bs{n} \right),
\\
\SPH{P}^{\partial \Omega}_t = \sum_{i \in \Omega} \sum_{j \in \Omega} \frac{m_i}{\rho_i} \frac{m_j}{\rho_j} \left(p_j \bs{u}_i \cdot \bs{t} + p_i \bs{u}_{j} \cdot \bs{t} \right)  \left( \Grad{W}_{ij'} \cdot \bs{t} \right).
\end{align}
Both projections can be split into two sums, conveniently swapping the indexes, in such a way that
\begin{align}
\SPH{P}^{\partial \Omega}_n = \sum_{i \in \Omega} \sum_{j \in \Omega} \frac{m_i}{\rho_i} \frac{m_j}{\rho_j} \left(
    \left(p_j \bs{u}_i \cdot \bs{n} \right)\left( \Grad{W}_{ij'} \cdot \bs{n} \right)
    -
    \left( p_j \bs{u}_{i} \cdot \bs{n} \right) \left( \Grad{W}_{j'i} \cdot \bs{n} \right)
\right),
\\
\SPH{P}^{\partial \Omega}_t = \sum_{i \in \Omega} \sum_{j \in \Omega} \frac{m_i}{\rho_i} \frac{m_j}{\rho_j} \left(
    \left(p_j \bs{u}_i \cdot \bs{t} \right)\left( \Grad{W}_{ij'} \cdot \bs{t} \right)
    +
    \left( p_j \bs{u}_{i} \cdot \bs{t} \right) \left( \Grad{W}_{j'i} \cdot \bs{t} \right)
\right).
\end{align}
However, due to the symmetry, it can be asserted that $\Grad{W}_{ij'} \cdot \bs{n} = \Grad{W}_{j'i} \cdot \bs{n}$.
Likewise, $\Grad{W}_{ij'} \cdot \bs{t} = - \Grad{W}_{j'i} \cdot \bs{t}$.
Consequently, $\SPH{P}^{\partial \Omega}_n = 0$ and $\SPH{P}^{\partial \Omega}_t = 0$. This fact also implies that the contribution from the binormal direction is also zero, which completes the proof.

\section{Wall force computation}
\label{s:mom_cons}
To compute the force exerted on the walls by the fluid, and vice-versa, momentum conservation is applied.

On one hand, the pressure force exerted by the fluid upon the walls can be expressed as
\begin{equation}
\bs{F}_\text{fw} = \sum_{j \in \partial \Omega} P_j \bs{n_j} s_j,
\end{equation}
where the value of the pressure at wall section $j$, $P_j$, is not known.
This would be the physical value of the wall pressure at $j$, different
from the $p_j$ used before, which is to be taken as a boundary condition of our problem.
However, the total pressure force exerted by the walls on the fluid is found
by using the pressure gradient of Eq. \eqref{eq:energy:homogeneous:gradp_bi}
in the momentum equation, Eq.~\eqref{eq:gov_equations:mom_cons}:
\begin{equation}
\bs{F}_\text{wf} =
- \sum_{i \in \Omega} 
\frac{m_i}{\rho_i} \sum_{j \in \partial \Omega} 2 p_i \bs{n_j} W_{ij} s_j.
\end{equation}

By imposing action and reaction, $\bs{F}_\text{fw} = -\bs{F}_\text{wf}$, we obtain a definition of the pressure at the boundary,
\begin{equation}
\label{capital_j_def}
P_j = 2 \sum_{i \in \Omega} p_i W_{ij} \frac{m_i}{\rho_i}.
\end{equation}
This result is precisely the SPH convolution of the fluid pressure for planar walls.
The pressure force on each $j$-th boundary element is therefore obtained as
$\bs{f}_j = P_j \bs{n_j} s_j$.
%


\end{document}